\documentclass[preprint]{jfm}

\usepackage{natbib}
\usepackage{graphicx}
\usepackage{amsmath}
\usepackage{amsfonts}
\usepackage{amssymb}
\usepackage{float}
\usepackage{bm}
\usepackage{color}

\newcommand{\Rep}{\mbox{\textit{Re}}_\mathrm{p}}
\newcommand{\Rel}{\mbox{\textit{Re}}_\lambda}
\renewcommand{\Re}{\mbox{\textit{Re}}}
\newcommand{\St}{\mbox{\textit{St}}}

\begin{document}

\title[Effect of turbulent fluctuations on the drag and lift forces on
  a sphere]{Effect of turbulent fluctuations on the drag and lift
  forces on a towed sphere and its boundary layer}

\author[H. Homann, J. Bec, R. Grauer]{H\ls O\ls L\ls G\ls E\ls R \ns
  H\ls O\ls M\ls A\ls N\ls N$^1$\footnote{Email address for
    correspondence: holger@oca.eu},  J\ls {\'E}\ls R\ls {\'E}\ls M\ls I\ls E
\ns B\ls E\ls C$^1$, \and R\ls A\ls I\ls N\ls E\ls R \ns G\ls R\ls A\ls U\ls E\ls R$^2$}

\affiliation{$^1$Laboratoire J.-L.\ Lagrange UMR\ls 7293, Universit\'e de Nice-Sophia Antipolis, CNRS,
  Observatoire de la C\^ote d'Azur,  BP4228, 06304 Nice, France.\\
  $^1$Theoretische Physik I, Ruhr-Universit\"at, 44780 Bochum, Germany.}

\maketitle

\begin{abstract}
  The impact of turbulent fluctuations on the forces exerted by a
  fluid on a towed spherical particle is investigated by means of
  high-resolution direct numerical simulations. The measurements are
  carried out using a novel scheme to integrate the two-way coupling
  between the particle and the incompressible surrounding fluid flow
  maintained in a high-Reynolds-number turbulent regime. The main idea
  consists in combining a Fourier pseudo-spectral method for the fluid
  with an immersed-boundary technique to impose the no-slip boundary
  condition on the surface of the particle. This scheme is shown to
  converge as the power 3/2 of the spatial resolution. This behaviour
  is explained by the $L_2$ convergence of the Fourier representation
  of a velocity field displaying discontinuities of its
  derivative. Benchmarking of the code is performed by measuring the
  drag and lift coefficients and the torque-free rotation rate of a
  spherical particle in various configurations of an upstream-laminar
  carrier flow. Such studies show a good agreement with experimental
  and numerical measurements from other groups. A study of the
  turbulent wake downstream the sphere is also reported. The mean
  velocity deficit is shown to behave as the inverse of the distance
  from the particle, as predicted from classical similarity
  analysis. This law is reinterpreted in terms of the principle of
  ``permanence of large eddies'' that relates infrared asymptotic
  self-similarity to the law of decay of energy in homogeneous
  turbulence.
  
  The developed method is then used to attack the problem of an
  upstream flow that is in a developed turbulent regime. It is shown
  that the average drag force increases as a function of the turbulent
  intensity and the particle Reynolds number. This increase is
  significantly larger than predicted by standard drag correlations
  based on laminar upstream flows. It is found that the relevant
  parameter is the ratio of the viscous boundary layer thickness to
  the dissipation scale of the ambient turbulent flow. The drag
  enhancement can be motivated by the modification of the mean
  velocity and pressure profile around the sphere by small scale
  turbulent fluctuations. It is demonstrated that the variance of the
  drag force fluctuations can be modelled by means of standard drag
  correlations. Temporal correlations of the drag and lift forces are
  also presented.
\end{abstract}

\section{Introduction}
\label{sec:intro}

The dynamics of suspended particles whose size is larger than the
smallest active scales of the carrier turbulent flow is important for
the understanding of many natural and industrial applications. Let us
mention two concrete problems borrowed from the study of planet
formation and of rain processes. In the first situation, the role of
turbulence is still unsettled in the dynamical mechanisms that take
place during the origination of planets in a proto-planetary disk
\citep[see, e.g.,][]{depater-lissauer:2001}. A key issue is to
understand whether or not the disk turbulence can generate an increase
of the rate at which a large body can accrete dust. In other terms, it
requires identifying the turbulent fluctuation mechanisms that lead to
a snowball effect where the larger bodies have a growth speed all the
more fast as their size increases. The same kind of question arises in
the precipitation process when a large raindrop falls under the effect
of gravity and collides with much smaller droplets that are
transported by the turbulent airflow in the cloud
\citep[see][]{shaw:2003}. For this problem, another question is to
know the efficiency of wet deposition (or scavenging) of particle
pollutants that are present in the atmosphere \citep[see,
e.g.,][]{seinfeld-pandis:1998}. These various phenomena are still
poorly understood and are subject to empirical ad-hoc modelling. Such
fundamental open questions, which are taken from natural sciences but
also arise in a number of industrial settings, reflect an inadequate
current level of modelling for large particles in turbulent flows.

Current approaches for the dynamics of finite-size particles are
commonly assuming that the particles are infinitely small (point
particles) and that the Reynolds number associated to their
interaction with the flow vanishes, so that their dynamics can be
obtained by matching the solution to the Stokes equation in their
vicinity to an unperturbed fluid velocity. Recent experimental work
showed that the dynamics of finite-size particles is not well
described by such models \citep[see][]{qureshi-bourgoin-etal:2007,
  xu-bodenschatz-etal:2008b, volk-calzavarini-etal:2010}. Significant
size effects have been for instance measured in the acceleration
statistics of such particles. Despite the fact that the flatness of
the acceleration distribution depends only weakly on the diameter of
the particle, its variance strongly decreases with diameter and
temporal correlations of the acceleration vary with the particle size
in a manner that cannot be predicted by standard point-particle
models. Improved models accounting for the so-called ``Fax\'{e}n
corrections'' (spatial averages of fluid velocities and accelerations
over the volume of the particle) are in much better agreement with
such qualitative trends but hardly reproduce quantitatively the
mentioned experimental data
\citep[see][]{calzavarini-volk-etal:2009,calzavarini-volk-etal:2011}.

A key ingredient of the finite-size models mentioned above are drag
correlations, which determine the forces on a sphere based on its
velocity difference with the fluid. All those drag correlations are
empirical formulas fitting experimental and numerical measurements of
the drag experienced by a sphere in a \emph{laminar} upstream
flow. However, the question whether or not turbulent fluctuations
eventually present in the carrier flow modify those drag correlations
is an open issue, which has been under discussion for more than forty
years. Several groups claim that turbulence increases the drag on a
sphere \citep[see][]{anderson-ulherr:1977,Brucato1998} while others
observe no significant modifications of the drag forces
\citep[see][]{warnica-renksizbulut-etal:1995a,Bagchi2003,Kim2011}. See
\cite{balachandar-eaton:2010} for a review. Recently,
\cite{luo-wang-etal:2010} have shown that a particle in an oscillating
flow experiences an increased drag force. It is difficult both for
numerical and for experimental measurements to reliably measure an
averaged force on the particle due to the statistical fluctuations
introduced by the turbulence.

Another challenging requirement to improve models for the dynamics of
finite-size particles in turbulent flows is to get a precise
understanding on how the carrier flow is modified in the vicinity of
the particle in situations where such effects occurs on scales which
are comparable to the active scales of the surrounding turbulence.
For instance, it is clear that much work is still needed in order to
get a better handling on how the kinetic turbulent energy, the
dissipation rate, as well as other standard statistical properties of
turbulence, varies in the vicinity of the particle. Direct numerical
simulations appear to be valuable candidates as they give a unique
global and instantaneous access to such quantities.  However, the
numerical problem is challenging for several reasons: \textit{(i)}\/
the numerical approach has to resolve all active scales of a fully
developed turbulent flow, which span several decades; \textit{(ii)}\/
at small scales the boundary layer around the particle is of great
importance and has to be well resolved; \textit{(iii)}\/ if one
considers a moving particle the algorithm has to follow the
displacement of the particle accurately and efficiently;
\textit{(iv)}\/ for a sufficient separation of the different scales
involved (forcing scales, inertial range of scales, dissipative
scales) the algorithm has to parallelize efficiently on massively
parallel computers.

Several different strategies have been recently proposed in
order to solve the incompressible Navier--Stokes equations together
with no-slip boundary conditions at the surface of a spherical
particle in a turbulent environment.  All of them make use
of a homogeneous grid covering the complete physical domain. 

``Physalis'', which was developed by \cite{prosperetti-oguz:2001} and
extended and used by \cite{naso-prosperetti:2010}, makes use of known
analytic solutions to the Stokes equation for the boundary layer in
the vicinity of the particle, in order to describe the velocity at the
nearest surface points of the grid. Iteratively, the analytic solution
and the solution on the homogeneous grid are matched. An advantage of
this strategy is that the no-slip condition is imposed in an exact
manner. However, improvement are required in order to achieve
very-large Reynolds number carrier flows \citep[the value $\Rel = 20$
of the Taylor-microscale Reynolds number was chosen
in][]{naso-prosperetti:2010}.  Another recently developed approach
consists in using and matching two different grids. A spherical one
encloses the particle surface and is designed to resolve the boundary
layer and a second homogeneous grid covers the entire domain. This
``Overset Grid'' technique proposed by \cite{burton-eaton:2005} uses a
third-order interpolation to transfer values from one grid to the
other in order to couple the two solutions. This method was shown to
reproduce very accurately experimental measurements of the drag
coefficient at a particle Reynolds number of $\Rep = 20$. In turbulent
settings, this method was used with a Reynolds number $\Rel =
32$. Finally, we note that \cite{Yeo-dong-etal:2010} uses a force
coupling method based on a low-order, finite force multipole
representation for the influence of the particles in the flow. This
efficient approach is limited to rather low particle Reynolds number
of less than 20. Most of these approaches are at the moment limited to
rather small carrier-flow Reynolds numbers. We propose here to use a
method that allows one to reach much higher Reynolds numbers. It
follows the approach of \cite{goldstein1993modeling} who used an
immersed boundary approach together with a spectral solver. However,
as \cite{mohd1997simulations}, we combine here a spectral method with
direct forcing instead of a feedback scheme \`a la
\cite{peskin1977numerical}.  Note that such a direct-forcing approach
has recently been used by \cite{uhlmann:2005} together with a
finite-difference method.

This paper is organised as follows. In \S~\ref{sec:method} we present
in details the proposed method, together with some considerations on
its convergence. Then, in \S~\ref{sec:benchmarks}, we report
measurements in a laminar upstream flow. The drag and lift
coefficients are measured in various settings and compared to other
numerical and experimental works. We also focus on the wake properties
and characterize the decay of turbulent fluctuations in the particle
wake. In \S~\ref{sec:turb} we turn to a particle in a fully-developed
stationary turbulent upstream flow and study this case varying the
turbulent intensity. The effect of fluid velocity fluctuations on the
drag and lift forces are discussed and interpreted in terms of the
modification of the particle boundary layer.  Finally, in
\S~\ref{sec:conclusion}, we draw some concluding remarks and suggest
open questions.

\section{Description of the numerical method}
\label{sec:method}

The idea of our approach is to combine a standard
pseudo-Fourier-spectral with a penalty method. The former is well
adapted to incompressible homogeneous and isotropic turbulence, has a
high degree of accuracy, and has shown good performances on massively
parallel supercomputers. No-slip boundary conditions are implemented
by an immersed boundary and penalisation strategy. The proposed method
is a modification of that used by \cite{homann-bec:2010} for
investigating the dynamics of neutrally buoyant particles suspended in
turbulent flows. The former version of the method was well-adapted to
particles with a small velocity difference with the fluid but happened
to show some limitations when the slip is too large. We present here a
modification of this method that reduces such shortcomings.

The dynamics of an incompressible flow in which is embedded a
spherical particle is given by the Navier-Stokes equations
\begin{equation}
  \label{eq:navier-stokes}
  \partial_t {\bm u} = \mathcal{L}(\bm u) = - \bm u\cdot\nabla\bm u -
  \nabla p +\nu \nabla^2 {\bm u} + \bm{f}_e, \quad\nabla \cdot {\bm
    u} =  0,\quad\mbox{for } |\bm x - \bm X_p(t)| > d/2
\end{equation}
together with the no-slip boundary condition on the surface of the
particle
\begin{equation}
  \label{eq:bc}
  {\bm u}(\bm x, t) = V_\mathrm{p}(t) + \bm\omega_\mathrm{p} (t) \times [\bm x - \bm X_\mathrm{p}
  (t)] \quad\mbox{for }  |\bm x - \bm X_p(t)| = d/2.
\end{equation}
$\bm f_e$ denotes here external forces that are exerted on the flow,
$d$ is the particle diametre, $X _\mathrm{p}$, $\bm V_\mathrm{p}$, and
$\bm\omega _\mathrm{p}$ are the particle position, translational
velocity, and angular velocity, respectively. The particle dynamics is
obtained by solving Newton's equations where the force exerted by the
fluid is obtained by integrating on the surface of the particle the
contribution from pressure and viscous stress to the fluid stress
tensor.

The numerical method presented here consists in integrating in a
rectangular periodic domain the Navier--Stokes equation
(\ref{eq:navier-stokes}) by using a strongly-stable, low-storage
Runge-Kutta scheme of third order that was introduced by
\cite{shu-osher:1988}. The spatial derivatives are evaluated in
Fourier space and the nonlinear term is computed with a
pseudo-spectral method using the standard dealiasing 2/3 rule
\citep[see, e.g.,][]{patterson-orszag:1970}.  The pressure term is
obtained by solving the Poisson equation directly in Fourier
space. The viscous term is treated implicitly via an exponential
factor.

\subsection{The penalty force}
To impose the no-slip boundary conditions (\ref{eq:bc}) at the surface
of the particle, we make use of an immersed boundary technique. It
consists in introducing in the right-hand side of the Navier-Stokes
equation (\ref{eq:navier-stokes}) a penalty force $\bm f_ b(\bm x,
t)$, which acts as a Lagrange multiplier associated to the constraint
defined by the boundary condition (\ref{eq:bc}). The full problem
(\ref{eq:navier-stokes})-(\ref{eq:bc}) can then be rewritten as
\begin{equation}
  \label{eq:ib}
  \partial_t {\bm u} = \mathcal{L}(\bm u) +\bm f_b, \quad\nabla \cdot
  {\bm u} = 0,
\end{equation}
that is satisfied in the whole domain. The core of such a method is to
find the best strategy to estimate $\bm f_b$. In principle the penalty
force should vanish outside the particle and be such that $\bm u$ is
equal to a solid motion inside it.  The Fourier representation of the
velocity field with a finite number of modes implies that these two
conditions cannot be satisfied in an exact manner. An idea could be to
find at each time step $\bm f_b$ with a minimal norm outside the
particle and such that the norm of the difference between $\bm u$ and
the particle velocity is minimal inside the particle.  This can be
done by standard minimisation techniques but such a strategy would
introduce a very large number of operations and result in a
cost-inefficient algorithm.

Efficient strategies consist in guessing acceptable values of the
field $\bm f_b$. We here make use of the ``direct-forcing'' method
used for instance by \cite{mohd1997simulations} and
\cite{fadlun-verzicco-etal-2000}.  To explain this method, consider
the simple Euler time discretisation of (\ref{eq:ib})
\begin{equation}
  \label{eq:euler}
  \bm u^{n+1} = \bm u^n + \Delta t \, \left[\mathcal{L}(\bm u^n) +
    \bm f_b^{n}\right].
\end{equation}
To impose the solid motion inside the particle, one writes
\begin{equation}
  \label{eq:directForcing}
  \bm f_b^{n} = -\chi_p(\bm x)\,\left[\mathcal{L}(\bm u^{n}) + (\bm
    u^n-\bm V_\mathrm{p}^{n+1}- \bm\omega_\mathrm{p}^{n+1} \times
    [\bm x - \bm X_\mathrm{p}^{n+1}])/\Delta t\right], 
\end{equation}
where $\chi_p(\bm x)$ denotes the characteristic function of the
particle: $\chi_p(\bm x) = 1$ if $|\bm x-\bm X_\mathrm{p}^{n+1}|<d/2$
and 0 otherwise.  Other similar approaches consist in imposing the
boundary conditions by means of a Darcy term, so that $\bm f_b =
-\alpha \, \chi_p(\bm x)\,(\bm{u} -\bm V_\mathrm{p}-
\bm\omega_\mathrm{p} \times [\bm x - \bm X_\mathrm{p}]))$
\citep[see][]{angot-bruneau-etal:1999}.  The particle volume is then
seen as a porous media where a large value of the parameter $\alpha$
ensures the no-slip boundary conditions but results in choosing a
sufficiently small value of the time step. The advantage of direct
forcing is that it consists in combining the value of the parameter
$\alpha$ that is optimal for a given choice of the time step $\Delta
t$, together with a compensation of the overall evolution of the
velocity field inside the particle due to $\mathcal{L}$.

The expression (\ref{eq:directForcing}) for $\bm{f}_b$ is not exact as
the position of the grid points do not coincide with the immersed
boundary. An interpolation procedure is thus needed and we use the
volume-fraction scheme proposed by \cite{fadlun-verzicco-etal-2000}
and \cite{pasquetti-bwemba-etal:2008}, which consists in averaging the
characteristic function over two grid cell:
\begin{equation}
  \bar \chi_p (\bm x) =\frac{1}{(2h)^3}\int_{-h}^{h}\int_{-h}^{h}\int_{-h}^{h}
  \chi_p(\bm x + \bm y) \,\mathrm{d}^3y.
\end{equation}
This representation has two advantages: it reduces Gibbs oscillations
in the vicinity of the boundary and it allows for a continuous
displacement of the particle.

\subsection{The pressure predictor}
\label{sec:pressurePredictor}
Another issue arises when combining a spectral with a penalty
method. Although applying the force $\bm{f}_b$ to the velocity field
imposes the boundary conditions precisely, the divergence-freeness of
$\bm u$ requires a subsequent projection of the velocity field onto
its solenoidal part which violates the formally correct boundary
conditions. A solution to this conflict is to solve the Poisson
equation for pressure with modified boundary conditions, as for
example proposed by \cite{taira-colonius:2007} in combination with a
finite-difference scheme. However, this is not possible with a
Fourier-spectral method as in this case the Poisson equation is solved
with generic periodic boundary conditions.

We attenuate this intrinsic problem of Fourier-spectral approaches by
using a predictor for the pressure gradient in (\ref{eq:ib}), as
considered by \cite{brown-cortez-etal:2001}. We will now explain
ourprocedure. The usual strategy in spectral codes without immersed
boundaries consists in advancing first the velocity without taking
into account the pressure gradient in (\ref{eq:navier-stokes}) and
then to project the resulting field onto its solenoidal part.  This is
equivalent to subtracting the gradient of the pressure. Such a
splitting of the integration is called a fractional-step method.  Now,
if the domain contains immersed boundaries with no-slip conditions,
the penalty force $\bm f_b$ has to be additionally applied. This
operation does not commute with the projection on divergence-free
vector fields. If the penalisation is done after imposing
incompressibility, the resulting field is compressible. Conversely if
it is done before, the boundary condition is not well satisfied.  We
attenuate this conflict by adding a predictor for the pressure
gradient. For this we use the value $\nabla p^{n}$ of the pressure
gradient at the last time step in the right-hand side of
Navier--Stokes equation. We then apply the penalty, so that the
resulting intermediate velocity field $\tilde{u}^{n+1}$ is very close to
divergence-free. Finally, $u^{n+1}$ is obtained by projection on
incompressible vector fields and the predictor is updated according to
\begin{equation}
  p^{n+1}=p^{n}+\Phi^n,\quad\mbox{where } \nabla^2 \Phi^n = \nabla\cdot
  {\tilde{u}}^{n+1}/\Delta t.
\end{equation}
This operations are of course done at each sub-timestep of the
third-order Runge--Kutta temporal scheme.

\begin{figure}
  \begin{center}
    \includegraphics[width=0.51\textwidth]{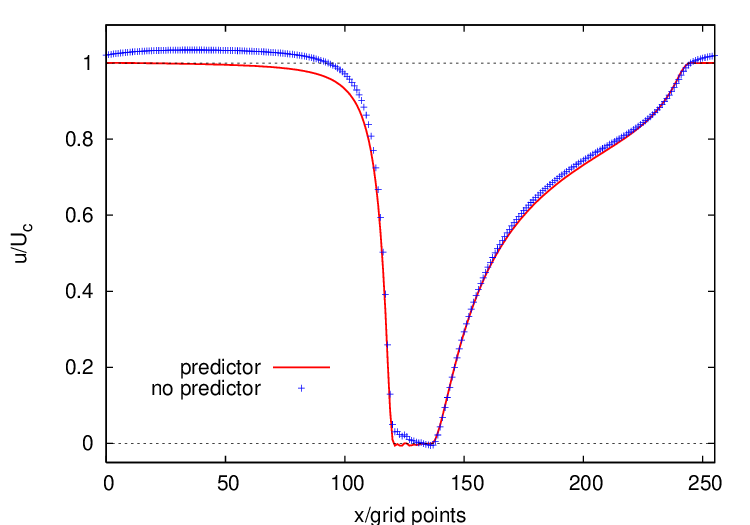}
    \hspace{-20pt}
    \includegraphics[width=0.51\textwidth]{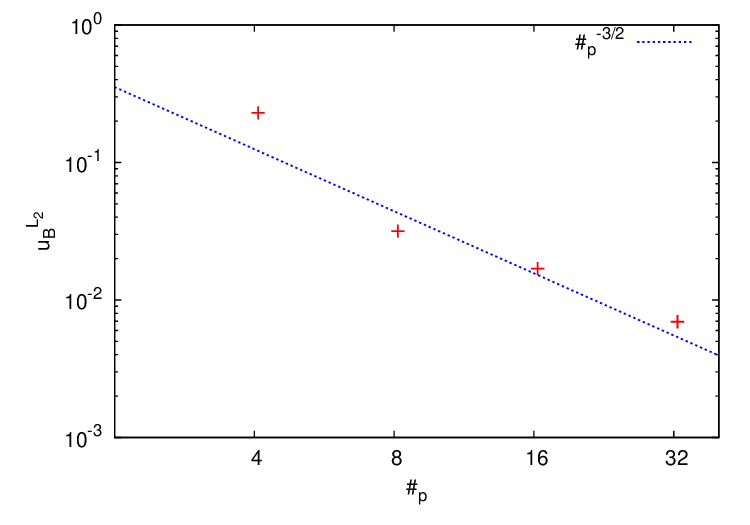} 
  \end{center}
  \caption{\label{fig:predictor_comp}(Left) Profile of the velocity
    (normalised to the inflow speed) in the direction of the mean flow
    on the axis of symmetry. The solid line corresponds to a
    computation with predictor $p^{n-1}$ while the points correspond
    to a computation without the predictor. (Right) $L_2$ norm of the
    velocity inside the particle as a function of the number of grid
    points on its diameter}
\end{figure}

The importance of the predictor for a Fourier-spectral scheme is
evident from Fig.~\ref{fig:predictor_comp} where the streamwise
velocity profile on a line passing through the particle is shown. The
flow conditions correspond to a numerical wind tunnel experiment where
a fixed spherical particle is embedded in a homogeneous flow. The
velocity fluctuations introduced by the wake of the particle are
damped in a zone at the right boundary by an additional application of
the penalty method. The accuracy of the upstream boundary condition at
the particle surface is strongly affected by the pressure
predictor. With it we approach better the no-slip condition at the
surface while the fluid significantly enters the particle
without. Also the smoothing of the velocity fluctuations at the exit
of the wind tunnel is much more efficient with the pressure predictor.

A measure for the accuracy of the boundary conditions is given by the
$L_2$-norm
\begin{equation}
\label{eq:l2norm}
  u_B^{L_2}=\left(\int_B |\bm{u}(x)|^2 dx\right)^{1/2}
\end{equation}
of the velocity inside the fixed spherical particle. This norm is
represented in Fig.~\ref{fig:predictor_comp} (Right) for a flow with a
particle Reynolds number $\Rep = 20$ and as a function of the number
of grid-points $\#_p$ along the diameter of the sphere. This number
quantifies the resolution of the geometry of the particle by the
underlying homogeneous grid. The norm $u_B^{L_2}$ shows a power-law
decrease with an exponent $-3/2$. This value can be understood by
remembering that the fluid velocity field is continous but not
differentiable at the surface of the particle. In a spectral scheme,
fields are represented by truncated series of $K$ modes.  One can then
easily show that the singular behavior of $\bm u$ at the surface leads
to $u_B^{L_2} \propto K^{-3/2}$.

\subsection{The forces on the sphere}
The force that is exerted by the fluid on the particle is given by the
integral over the surface of the particle of the fluid stress tensor
\begin{equation}
  \label{eq:forceParticle}
  \bm{F} = \int_{\Omega_\mathrm{p}(t)} \nabla \cdot \mathbb{T} \,\,
  \mathrm{d}^3x = \int_{\partial \Omega_\mathrm{p}(t)} \mathbb{T}
  \cdot\mathrm{d}\bm S,\qquad \mbox{where} \quad 
  \mathbb{T}=-p\bm{I}_3+\nu\, (\nabla \bm{u}+{\nabla \bm{u}}^T).
\end{equation}
The torque on the particle is given by
\begin{equation}
  \label{eq:torque}
  \bm{T}=\int_{\partial
    \Omega_\mathrm{p}(t)}\bm{n}\times(\mathbb{T}\cdot \mathrm{d}\bm
  S)
\end{equation}
where $\bm{n}$ is the unit normal vector of the particle surface.  The total force $\bm F$
can easily be computed from the penalty force $\bm f_b$ because
\begin{equation*}
  \int_{\Omega_\mathrm{p}(t)} \nabla \cdot \mathbb{T} \,\,
  \mathrm{d}^3x = \int_{\Omega_\mathrm{p}(t)}  \bm f_b\,
  \mathrm{d}^3x.
\end{equation*}
In order to distinguish the pressure and viscous terms appearing in
the stress tensor (\ref{eq:forceParticle}), our strategy consists in
computing the pressure contribution by constructing a homogeneous grid
of points at the sphere surface. Then, we integrate the directed
pressure $p\bm{n}$, $\bm{n}$ being the normal vector field at the
surface of the particle, over this grid. The viscous part is obtained
by subtracting the pressure contribution from the total force.

\section{A particle in a laminar upstream flow}
\label{sec:benchmarks}
 
\begin{figure}
  \begin{center}
    \includegraphics[width=0.7\textwidth]{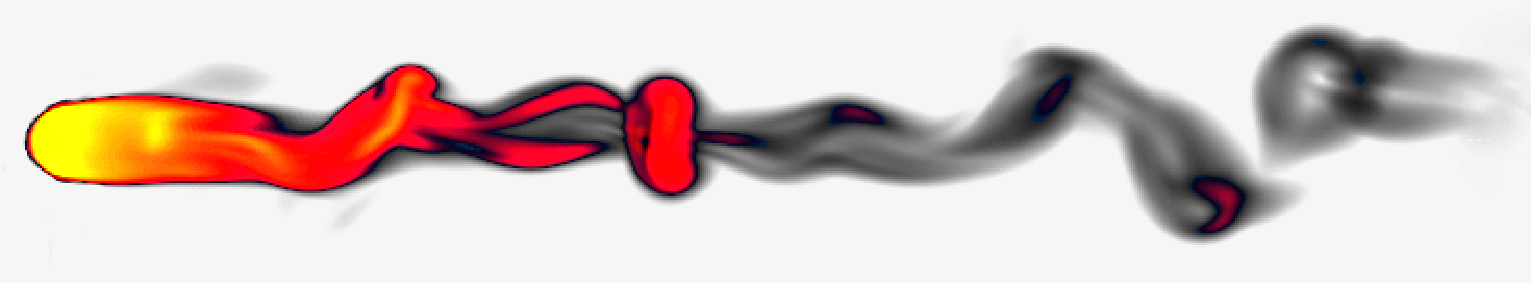} 
  \end{center}
  \caption{\label{fig:Re400_9000} Turbulent wake at $\Rep=400$ (volume
    rendering of high-vorticity isosurfaces).}
\end{figure}

We conduct here a numerical wind tunnel experiment, as already
described above. The sphere is fixed and the inflow velocity $U_c$ is
prescribed to be equal to one. The Reynolds number is varied by
changing either the particle diametre $d$ or the viscosity of the
fluid. The principle flow configuration is depicted in
Fig.~\ref{fig:Re400_9000} above the critical Reynolds number where a
non-stationary wake develops.
The goal of such settings is two-fold. First, it is used to perform
several benchmarkings of the proposed numerical method. Secondly, it
used to investigate the decay of the turbulent wake at moderate
Reynolds numbers for which we propose an approach in terms of the
``principle of permanence of large eddies''.

\subsection{Forces acting on the particle}
\begin{figure}
  \begin{center}
    \includegraphics[width=0.495\textwidth]{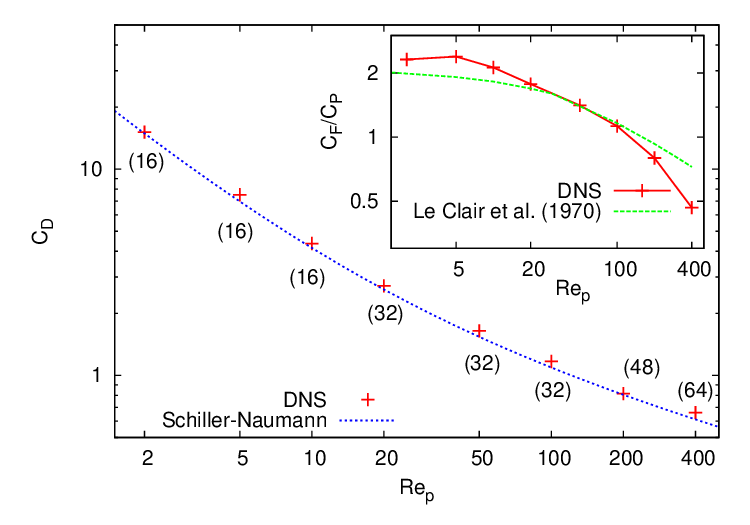}
    \includegraphics[width=0.495\textwidth]{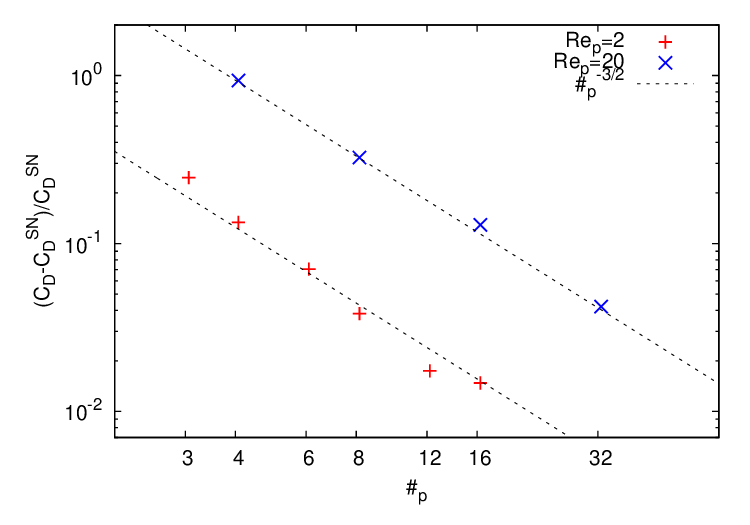}
  \end{center}
  \caption{\label{fig:re_drag_coeff} (Left) Drag-coefficients $C_D$ as
    a function of the particle Reynolds number; in parenthesis is
    given the number of internal grid points. Inset: ratio between the
    viscous contribution $C_F$ to the drag coefficient and the
    contribution $C_P$ from pressure. (Right) Convergence rate as a
    function of the number of grid points $\#_p$ across the sphere.}
\end{figure}

We first report measurements on the drag-coefficient
\begin{equation}
  C_D=\frac{8\,F_D}{\pi d^2\,\rho_\mathrm{f}U_c^2 },
\end{equation}
where $F_D$ denotes the total force exerted on the particle in the
stream-wise direction and $\rho_\mathrm{f}$ the fluid mass density
(set here to one). Figure \ref{fig:re_drag_coeff} shows the measured
coefficient. Data compare well to \cite{schiller-naumann:1933} empirical
formula
\begin{equation}
  \label{eq:schillerNaumann}
  C^{\mathrm{SN}}_D=\frac{24}{\Rep}(1+0.15\Rep^{0.687}).
\end{equation}
The numbers in parenthesis in Fig.~\ref{fig:re_drag_coeff} (Left)
denote the number of grid-points $\#_p$ across the sphere diametre $d$
and controls the resolution of the boundary layer around the
sphere. The thickness of this layer decreases as
$\Re^{-1/2}\,d$. Hence higher is the Reynolds number, more points
$\#_p$ are required. In order to check to performance of the scheme in
the case where the particle is moving, we changed the frame of
reference of this experiment and translated the particle in a flow at
rest. Also the zone where we remove the velocity fluctuations is moved
according to the particle position. The measured drag coefficients and
the velocity profile around the sphere are nearly indistinguishable
from the former wind tunnel experiments. We observe fluctuating
deviations from the values given in Fig.~\ref{fig:re_drag_coeff} of
less than one percent for Reynolds number up to 50 and less than five
percent up to 400. The inset of Fig.~\ref{fig:re_drag_coeff} (Left)
shows the ratio of the viscous $C_F$ to the pressure $C_P$
contribution to the complete drag $C_D = C_F+C_P$. The measured ratios
are in reasonable agreement with the results of
\cite{LeClair-hamielec-etal:1970}, who used a finite-difference
scheme. A possible explanation of the discrepancy could stem from the
method we use to separate the viscous and pressure
contributions. However, we stress again that such a distinction of these
two parts to the total force is not needed for the integration
of the particle dynamics and might only be of interest for analysis
purposes.

The convergence of the drag coefficient can be quantified by looking
at the relative error $(C_D-C^{\mathrm{SN}})/C^\mathrm{SN}$ as a
function of the internal points $\#_p$. From
Fig.~\ref{fig:re_drag_coeff} (Right) we observe that the coefficients
converge with an exponent of -3/2 for the two Reynolds numbers showed
there. This is the same rate of convergence as for the $L_2$ norm
$u_B^{L_2}$ of the velocity inside the particle (see previous
section), which suggests that the convergence rate of the proposed
scheme is determined by the convergence rate of the truncated
representation of the $C^0$-velocity field.

We next consider a non-homogeneous inflow with a constant gradient
$U_y = U_c + g\,(y-y_0)$, caracterized by the non-dimensional shear
rate $s={g\,d}/{U_c}$. This linear velocity profile creates a lift
force $F_L=\bm{F}\cdot\bm{e}_y$ on the particle, whose associated lift
coefficient is
\begin{equation}
  C_L=\frac{8\,F_L}{\pi d^2\rho_\mathrm{f}  U_c^2}.
\end{equation}
This coefficient have been measured by different groups:
\cite{dandy-dwyer:1990} used finite volumes, \cite{kurose-komori:1999}
finite differences, and \cite{bagchi-balachandar:2002} a
Chebyshev--Fourier method. All of them used a cylindrical mesh to
reach a higher resolution near the particle surface.
\begin{figure}
  \begin{center}
    \includegraphics[width=0.495\textwidth]{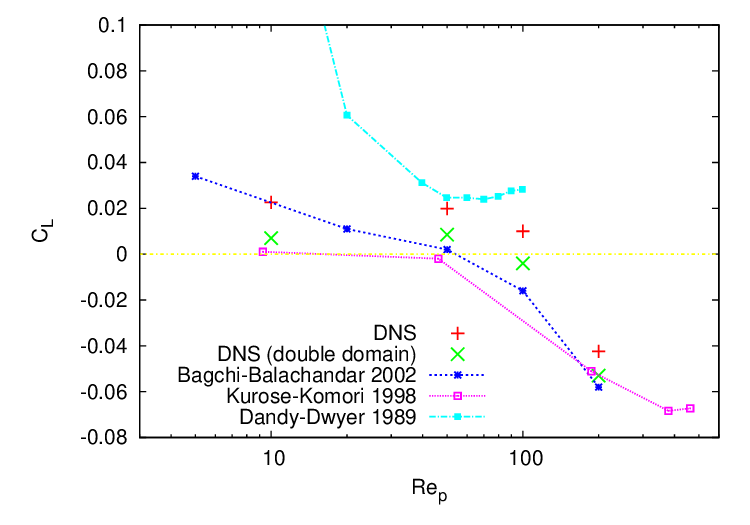} 
    \includegraphics[width=0.495\textwidth]{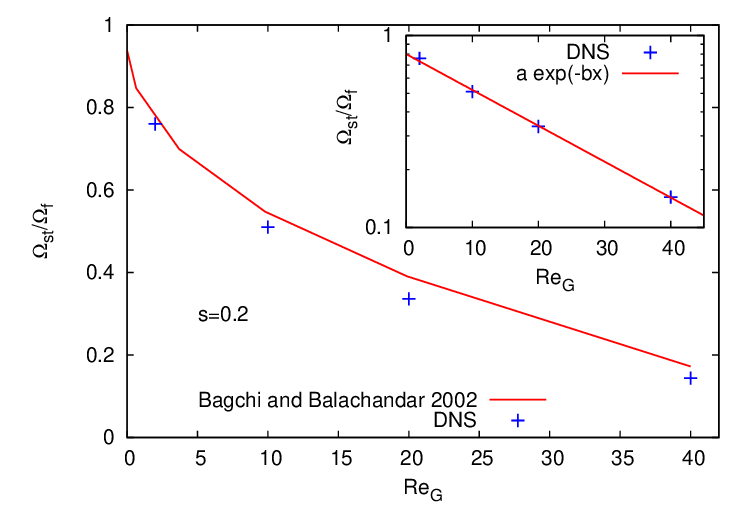}
  \end{center}
  \caption{\label{fig:lift_coeff_sphere} (Left) Lift-coefficients
    $C_L$ as a function of the particle Reynolds number for
    $s=0.2$. (Right) Stationary rotation rate (normalised to the
    ambient rotation rate $\Omega_f=g/2$) as a function of the
    particle Reynolds number defined with the mean fluid velocity
    gradient $\Re_\mathrm{G} = gd^2/\nu$. Inset: Logarithmic
    representation. For the exponential fit we used $a=0.8$ and
    $b=0.043$.}
\end{figure}
From Fig~\ref{fig:lift_coeff_sphere} (Left) one observes a quite large
spread of the lift coefficients measured by these three groups. This
is due to the sensitivity of the lift force on the stresses in the
flow and thus on the size of the computational domain. It has been
claimed by \cite{bagchi-balachandar:2002} that the box size in the
work of \cite{dandy-dwyer:1990} was too small, leading to an
overestimation of the lift force. We have computed the lift force for
two different box sizes ($128\!\times\!128\!\times\!256$ and
$256\!\times\!256\!\times\!512$) and observe a decreasing force with
increasing box-size. Our measurements are in good qualitative
agreement with the two most recent works and reproduce in particular
the transition to negative values for $\Rep \approx 60\!-\!80$. To
address in more details this question, larger simulations would be
required; this is beyond the scope of this work.

We have finally considered a freely rotating sphere in this shear
flow. After a relaxation period, the sphere torque-free rotation rate
attains a stationary value, which is shown in
Fig.~\ref{fig:lift_coeff_sphere} (Right) as a function of the shear
rate.  The measured values agree well with those of
\cite{bagchi-balachandar:2002}. In particular, the decay is slower
than the algebraic form proposed by \cite{lin-etal:1970}. As seen in
the inset, we actually observe that the rotation rate decreases
exponentially with the Reynolds number. However, this might again
result from a finite-box effect.

We have seen in this section that the drag and lift coefficients in a
laminar upstream flow are reasonably well captured by the proposed
numerical method. We will now turn to investigating the decay of the
turbulent wake. Before this let us do some comments on the effect of
Gibbs oscillations on the accuracy of the applied numerical
method. The representation of a discontinuous function with a finite
number $N/3$ of Fourier modes introduces oscillations in real
space. In our case the velocity gradient displays such oscillations
whose amplitude decreases as $1/(k_{max}\,x)$ where $x$ is the
distance to the particle surface and $k_{max}=(2\pi/L)\,N/3$. As we
will see in the next section the velocity gradient in the turbulent
wake of the particle decays as $x^{-3/2}$, that is faster than Gibbs
oscillations. This implies that there is a distance $x^*\simeq
C\,k_{max}^2$ beyond which the error takes over. This gives some
limitations in the use of this method to investigate the very-far
wake. Nevertheless, its quadratic dependence on $k_{max}$ makes $x^*$
grow fast as a function of the resolution. In addition we expect the
constant $C$ to increase with the particle Reynolds number as the
gradients get steeper. It is thus very likely that the proposed method
is adapted to study small-scale statistics in the turbulent wake for
the parameter range of our simulations. Also let us stress that in the
case when the outer flow is itself in a developed turbulent state, the
gradient has a typical amplitude so that Gibbs oscillations are always
sub-dominant at sufficiently far distances.

\subsection{Turbulent wake at moderate Reynolds numbers}
\label{sec:wake}

We now focus on the fluid flow velocity and measure the turbulent
velocity statistics in the wake of the particle. Most models for
turbulent flows past an obstacle rely on studying solutions to the
Reynolds-averaged equations for the stationary average $\bm U(\bm x) =
\langle \bm u(\bm x, t)\rangle$ velocity field. This equation involves
the Reynolds stress tensor $\tau_{ij}(\bm x) = \langle u'_i(\bm
x,t)\,u'_j(\bm x,t)\rangle$, where $\bm u' = \bm u -\bm U$ denotes
here the turbulent fluctuation. It is generally closed using an
eddy-viscosity assumption \citep[see, e.g.,][]{pope:2000}. When the
eddy viscosity $\nu_\mathrm{T}$ is assumed to be constant in space,
one obtains that the streamwise mean velocity $U(x,y,z) = \langle
u_x(\bm x, t)\rangle$ has the self-similar form \citep[see,
e.g.,][]{schlichting:1979}
\begin{equation}
  U(x,y,z) = U_c\left[ 1 - \frac{C_D \,d^2}{32\,\nu_\mathrm{T}x}
    \,\mathrm{e}^{-\frac{y^2+z^2}{r_{1/2}^2(x)}}\right],
  \mbox{ with } r_{1/2}(x) =
  d\,\sqrt{\frac{2\,U_c\,x}{\nu_\mathrm{T}\,\ln 2}}.
  \label{eq:selfsim_profile}
\end{equation}
$r_{1/2}$ is called the wake half-width.  Note that we assume in this
section that the particle is located at $x_0 = 0$.  Such a formula,
which is commonly used in industrial applications, implies that the
mean velocity deficit $\Delta U = 1 - U(x,0,0)/U_c$ behaves as
$x^{-1}$ at large distances. This law was observed experimentally
\citep[][]{wu-faeth:1994} and numerically
\citep[][]{bagchi-balachandar:2004}.

\begin{figure}
  \begin{center}
    \includegraphics[width=0.51\textwidth]{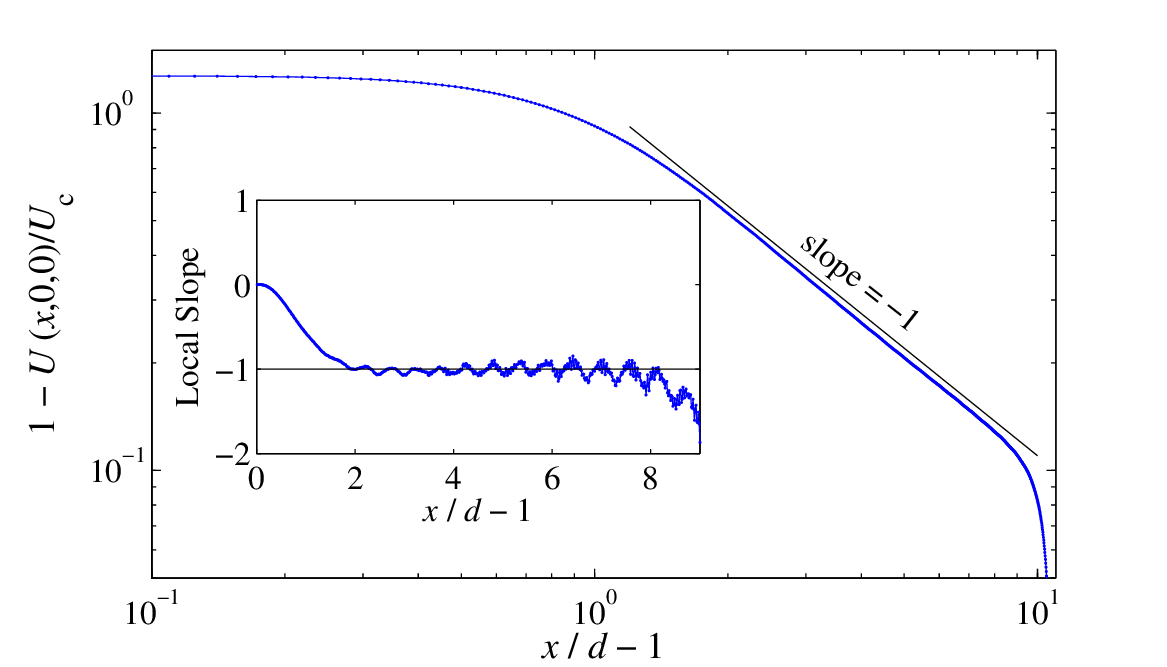}
    \hspace{-15pt}
    \includegraphics[width=0.51\textwidth]{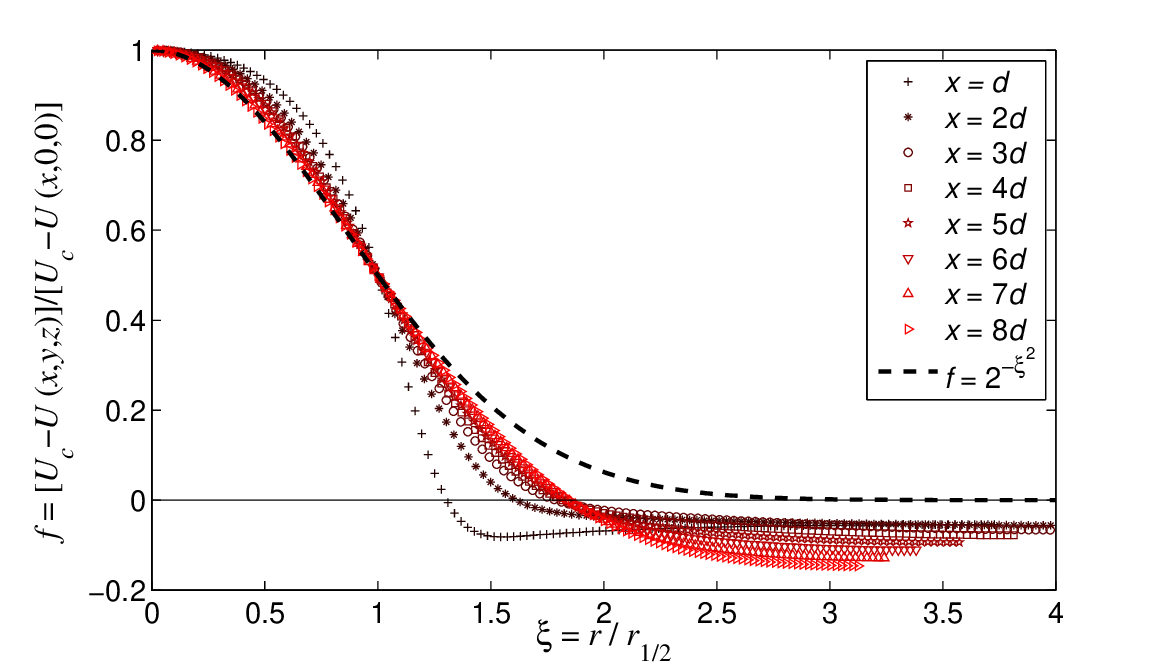}
  \end{center}
  \caption{\label{fig:U_decay} (Left) mean velocity deficit $\Delta U
    = 1 - U(x,0,0)/U_c$ in the axis of symmetry $y=0$ and $z=0$ of the
    wake as a function of the normalised streamwise distance
    $x/d$. The inset represents its logarithmic derivative
    $[\mathrm{d}\ln \Delta U]/[\mathrm{d}\ln (x/d-1)]$. (Right) mean
    velocity deficit profile $f$ as a function of the reduced spanwise
    distance $\xi = r/r_{1/2}$ for various values of $x/d$ as
    labelled; the bold dashed line represents the Gaussian functional
    form of $f$ obtained from the constant-eddy-viscosity similar
    profile \eqref{eq:selfsim_profile}.}
\end{figure}

An analysis of the turbulent wake from a direct simulation performed
at the moderate value of the particle Reynolds number $\Rep = 400$
shows that this similar eddy-viscosity theory rightly predicts the
decay of the mean velocity deficit. This is clear in
Fig.~\ref{fig:U_decay} (Left) where $\Delta U$ decreases like a power
law with exponent $-1$ over nearly half a decade. The oscillations
around the value $-1$ that we observe for the local slope in the inset
are a signature of the Gibbs phenomenon induced by the strip forcing
at $x\approx12d$.

However, as seen from the right-hand panel of Fig.~\ref{fig:U_decay},
the mean velocity deficit profile does not seem to display any
similarity. The half width $r_{1/2}$ was computed as the value of the
distance to the axisymmetric axis where the velocity deficit is equal
to $1/2$. One observes discrepancies for values of the reduced
variable $\xi = r/r_{1/2} = \sqrt{y^2+z^2}/r_{1/2}$ that are close to
$1/2$, to $3/2$, and for $\xi>2$. The deviation at large $\xi$'s might
be due to the periodicity of the spatial domain in the $y$ and $z$
directions.  However the deviations at moderate values are comparable
to those obtained experimentally by \cite{uberoi-freymuth:1970} and
\cite{wu-faeth:1994}.

In order to get a better handling on the observed $\propto x^{-1}$
behaviour, we relate it now to the decay of turbulent kinetic energy
in the wake of the particle. For this, we introduce an original
approach based on the ``principle of permanence of large eddies'' that
is known since the seminal work of \cite{kolmogorov:1941b} for the
time decay of homogenous turbulence. We refer here the reader to the
phenomenological presentation of this law made by \cite{frisch:1995}.
The principle of permanence of large eddies states that if the initial
turbulent velocity correlations have the form $\langle u(x+r) \,
u(x)\rangle \sim r^{-h}$ at large scales $r$, the decay behaviour of
all turbulent quantities is dictated by the value of the exponent
$h$. In particular, the integral scale is $L\sim t^{1/(1+h)}$ and the
kinetic turbulent energy behaves as $k\sim t^{-2h/(1+h)}$.  The main
principle used in this approach is the continuity of correlations at
$r=L$.

When considering the decay of wake turbulence, all the turbulent
quantities have to be of course evaluated in planes perpendicular to
the axisymmetric axis and time has to be replaced by $x/U_c$ by
invoking Taylor hypothesis.  Also, instead of using continuity at
separations equal to the correlation length, the main ingredient is
here that the spatial correlations of turbulent fluctuations have to
match the decay of the velocity deficit at very large transverse
distances $r$. The principle of permanence of large eddies then
implies that if the velocity deficit $\Delta U(r)$ at a distance $r$
from the axis of symmetry behaves as $r^{-h}$ in the developing wake,
then at any $x>0$ in the far wake, the large-scale correlations of the
turbulent fluctuations should remember this initial form and the
turbulent kinetic energy is expected to decay as $k \sim
x^{-2h/(1+h)}$.

\begin{figure}
  \begin{center}
    \includegraphics[width=0.8\textwidth]{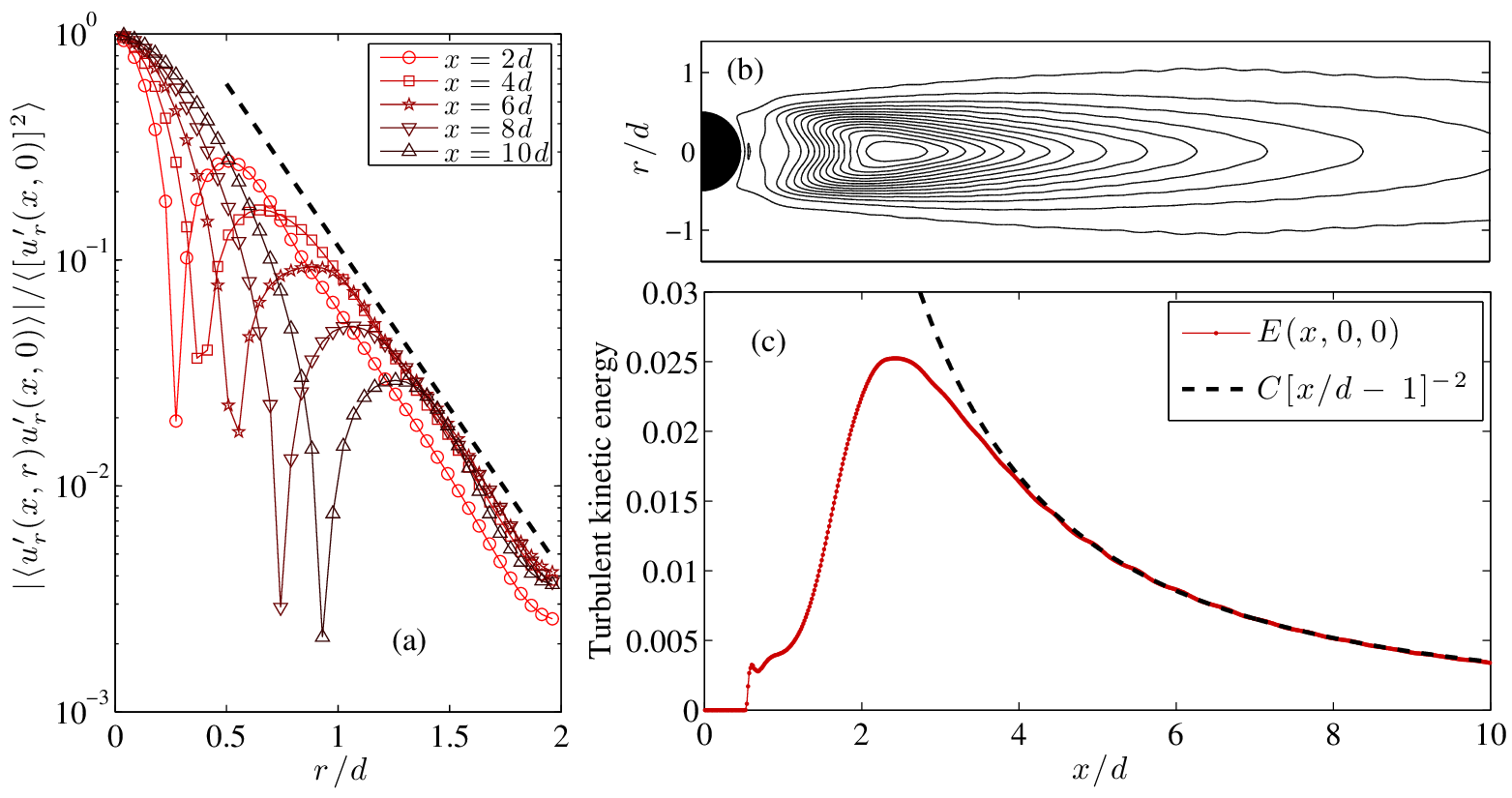}
  \end{center}
  \caption{\label{fig:correl_ur} (a) Correlation of the velocity
    radial fluctuation as a function of the distance $r$ to the
    axisymmetric axis for various values of the streamwise distance
    $x$ as labelled; the dashed line corresponds to an exponential
    decay. (b) Contour lines of the turbulent kinetic energy $k$ in
    the $(x,r)$ plane. (c) Profile of $k$ along the $r=0$ axis,
    clearly displaying a behaviour $\propto x^{-2}$}
\end{figure}

We now give an argument demonstrating that the velocity deficit decays
exponentially at large transverse distances $r$.  For that, we assume
that, to leading order, the axisymmetric mean velocity profile takes
for $r\gg d$ the separated form
\begin{equation}
  U_\parallel \simeq U_c+ \Phi_\parallel(r)\,\Psi_\parallel(x) \quad
  \mbox{and} \quad U_\perp \simeq \Phi_\perp(r)\,\Psi_\perp(x),
\label{eq:ansatz_separation}
\end{equation}
in the streamwise and transverse directions, respectively. We choose
here $\Phi_\parallel>0$ and $\Phi_\perp>0$, so that $\Psi_\parallel<0$
and $\Psi_\perp<0$. Incompressibility implies $\Phi_\parallel
\Psi_\parallel' + (1/r)\,(r\Phi_\perp)' \,\Psi_\perp = 0$, so that
\begin{equation}
(r\Phi_\perp)' = C_1\,r\,\Phi_\parallel \quad \mbox{and} \quad
\Psi_\parallel' = -C_1\,\Psi_\perp,
\label{eq:incompr}
\end{equation}
where $C_1$ is a constant, which by dimensional analysis should be
$\propto 1/d$. To leading order, the stationary Navier--Stokes
equation yields
\begin{equation}
 U_c\,  \Phi_\parallel\,\Psi_\parallel' \simeq -\partial_x p\quad
  \mbox{and} \quad U_c\, \Phi_\perp\,\Psi_\perp' \simeq -\partial_r p.
\label{eq:ns}
\end{equation}
The leading behaviour of the non-linear term is here given by the
advection of the perturbed velocity by the mean flow $U_c$. Note that
the Reynolds stress involves the square of the turbulent fluctuations
and is thus subleading when $r\gg d$. As the viscous terms become
negligible at large distances, advection is exactly compensated by the
pressure gradient. We now equate the second-order cross derivative
$\partial_x\partial_r p$ of pressure obtained from these two equations
to write $\Phi_\parallel'\Psi_\parallel' = \Phi_\perp\Psi_\perp''$,
leading to
\begin{equation}
\Phi_\parallel' = C_2\,\Phi_\perp \quad \mbox{and} \quad
\Psi_\perp'' = C_2\,\Psi_\parallel',
\label{eq:comp_ns}
\end{equation}
where again $C_2$ is a constant $\propto 1/d$. Put together, the
conditions (\ref{eq:incompr}) and (\ref{eq:comp_ns}) imply that
$r\Phi_\perp$ is a solution of the differential equation
\begin{equation}
(r\Phi_\perp')' = C_1\,C_2\,r\, \Phi_\perp, \quad \mbox{so that} \quad
\Phi_\perp \propto \frac{1}{\sqrt{r}}\,\mathrm{e}^{-C\,r/d}, \mbox{
  when } r\gg d
\label{eq:fin_res}
\end{equation}
with $C = d\,\sqrt{C_1\,C_2}$ a positive dimension-less constant that
is independent of $d$. This exponential decay of the mean velocity
deficit implies the same behavior for velocity correlations, as
confirmed numerically.  We indeed see in Fig.~\ref{fig:correl_ur}~(a)
that, for various values of $x$, the correlations of the radial
component of the fluctuating velocity collapse for large $r$ to an
exponential behaviour $\propto \exp (-r/r_0)$.  To our knowledge this
behaviour has never been observed before.

Turnig back to the law of decay, an exponential behaviour of
correlations corresponds to $h=\infty$. This implies that the
turbulent kinetic energy decays as $k\sim x^{-2}$, as confirmed
numerically in Fig.~\ref{fig:correl_ur}~(c).  Such a law of turbulence
decay can in turn be related to the downstream decrease of the
velocity deficit. The Reynolds-averaged Navier--Stokes equation can be
used to show that $\Delta U(x) = 1-U(x,0,0)/U_c$ approximately obeys
\begin{equation}
  U_c^2\, \partial_x \Delta U \sim \partial_x \tau_{xx}  + \partial_y \tau_{xy} + \partial_z
  \tau_{xz} \sim k / L \propto x^{-2} , \qquad \mbox{so that}\quad
  \Delta U \sim x^{-1}.
\end{equation}
$\bm\tau$ denotes here the Reynolds stress tensor. This finally
implies that the scaling of the velocity deficit is consistent with
the decay of turbulent kinetic energy. Also $h=\infty$ implies that
the integral scale remains constant ($L\sim x^0$). The kinetic energy
dissipation rate is thus decreasing as $\varepsilon\sim k^{3/2}/L\sim
x^{-3}$, so that the typical turbulent gradients behave as
$\partial_xu\sim (\varepsilon/\nu)^{1/2}\sim x^{-3/2}$.

We have seen in this section that the numerical method we are using is
well adapted to describe the turbulent wake downstream the
particle. This is for instance supported by the accuracy with which
the decay laws are measured. This can be understood by the fact that
Fourier-spectral methods are well adapted to simulations of developed
turbulence. To take advantage of this, we next turn to investigating
the influence of turbulent fluctuations that are possibly
present in the upstream flow.

\section{A particle in a turbulent upstream flow}

We now consider that the fluid flow surrounding the particle is forced
to sustain a turbulent state.  The simulations were set up in the
following way: first we precomputed a homogeneous, isotropic turbulent
flow in a periodic box without particle by solving
(\ref{eq:navier-stokes}) with $\Omega_p=\emptyset$. A statistically
stationary flow is reached by keeping constant the energy content in
the two lowest Fourier shells of the velocity field. On this turbulent
flow we superimposed a constant mean flow $\bm{U}_c=U_c\,\bm{e}_x$ by
keeping constant the real part of the zero Fourier mode. Afterwards, a
fixed sphere is placed into this fluctuating stream by means of the
penalty method discussed above. We thus solve (\ref{eq:navier-stokes})
with (\ref{eq:bc}) in a periodic domain while choosing its size
sufficiently large in the streamwise direction so that the sphere does
not interact with its own wake. The upstream flow seen by the sphere
can therefore be considered homogeneous and isotropic.  We have
performed four series of numerical simulations corresponding to four
different particle Reynolds numbers $\Rep$ (defined with the mean flow
$U_c$). From the range considered in previous section we choose
$\Rep=20$, $\Rep=100$, $\Rep=200$, and $\Rep=400$ in order to cover
three qualitatively different wake configurations. For $\Rep=20$ the
wake is intrinsically laminar without any recirculation region behind
the particle, for $\Rep=100$ and $\Rep=200$ a laminar recirculation
region develops and for $\Rep=400$ the wake becomes intrinsically
turbulent. The parameters of the simulations are given in
table~\ref{table}.

\begingroup
\begin{table*}
  \centering
  \begin{tabular}{ccccccccccccc}
    $\Re_p$ & $\Re$  &$u_\mathrm{rms}$&$\varepsilon$&$\nu$ & d& $\eta$&$\tau_\eta$&$L$   &$T_L$& $N^3$ \\
    \hline 
    20     & $0$   & $0$           &$1.7\cdot 10^{-4}$  &0.02& 0.4 &$-$    & $-$       &$-$  & $-$  &$1024\times 256^2$ \\
    20     & $0.31$& $0.05$        &$4.7\cdot 10^{-4}$  &0.02& 0.4 &(0.36) & (6.5)     &(0.15)&(3.7)&$1024\times 256^2$ \\
    20     & $0.94$& $0.064$       &$6.4\cdot 10^{-4}$  &0.02& 0.4 &(0.34) & (5.5)     &(0.32)&(5.4)&$1024\times 256^2$ \\
    20     & $3.2$ & $0.12$        &$2.64\cdot 10^{-3}$ &0.02& 0.4 &$0.24$ & 2.85      &0.56 & 4.9  &$1024\times 256^2$  \\
    20     & $23$  & $0.27$        &$1.23\cdot 10^{-2}$ &0.02& 0.4 &$0.16$ & 1.38      &1.67 & 6.1  &$1024\times 256^2$ \\
    20     & $39$  & $0.37$        &$2.40\cdot 10^{-2}$ &0.02& 0.4 &$0.135$& 0.91      &2.11 & 5.7  &$1024\times 256^2$ \\
    20     & $78$  & $0.57$        &$6.65\cdot 10^{-2}$ &0.02& 0.4 &$0.11$ & 0.55      &2.75 & 4.9  &$1024\times 256^2$ \\
    \hline 
    100    & 0    & 0              &$1.5\cdot 10^{-4}$  &0.006& 0.6 & $-$    & $-$      &$-$  & $-$  &$1024\times 256^2$ \\
    100    & 49.1 & 0.135          &$1.13\cdot 10^{-3}$ &0.006& 0.6 & 0.11   & 2.14     &2.18 & 16.1 &$1024\times 256^2$ \\
    100    & 140  & 0.28           &$7.3 \cdot 10^{-3}$ &0.006& 0.6 & 0.074  & 0.90     &3.01 & 10.7 &$1024\times 256^2$ \\
    100    & 288  & 0.58           &$6\cdot 10^{-2}  $  &0.006& 0.6 & 0.043  & 0.32     &2.93 & 4.97 &$1024\times 256^2$ \\
    \hline 
    200    & 0    & 0              &$5.4\cdot 10^{-5}$  &0.003& 0.6 &$-$     & $-$      &$-$  & $-$  &$2048\times 256^2$ \\
    200    & 130  & 0.135          &$8.5\cdot 10^{-4}$  &0.003& 0.6 &$0.075$ & $1.85$   &2.89 & 21.4 &$2048\times 256^2$ \\
    200    & 305  & 0.28           &$6.7\cdot 10^{-3}$  &0.003& 0.6 &$0.045$ & $0.67$   &3.28 & 11.7 &$2048\times 256^2$ \\
    200    & 426  & 0.4            &$2.0\cdot 10^{-2}$  &0.003& 0.6 &$0.034$ & $0.39$   &3.20 & 8.0  &$2048\times 256^2$ \\
    \hline 
    400    & 0    & 0              &$7.5\cdot 10^{-5}$  &0.002& 0.8 &$0.102$ & $5.2$    &$-$  & $-$  &$2048\times 256^2$ \\
    400    & 46.1 & 0.061          &$1.5\cdot 10^{-4}$  &0.002& 0.8 &$0.086$ & $3.7$    &1.51 & 24.81&$2048\times 256^2$ \\
    400    & 65.0 & 0.0714         &$2.0\cdot 10^{-4}$  &0.002& 0.8 &$0.079$ & $3.15$   &1.82 & 25.49&$2048\times 256^2$ \\
    400    & 203  & 0.139          &$9.2\cdot 10^{-4}$  &0.002& 0.8 &$0.054$ & $1.46$   &2.92 & 21.0 &$2048\times 256^2$ \\
    400    & 434  & 0.286          &$7.7\cdot 10^{-3}$  &0.002& 0.8 &$0.032$ & $0.51$   &3.04 & 10.63&$2048\times 256^2$ \\
    400    & 605  & 0.392          &$1.95\cdot 10^{-2}$ &0.002& 0.8 &$0.025$ & $0.32$   &3.09 & 7.88 &$2048\times 256^2$ \\
    400    & 927  & 0.598          &$6.9\cdot 10^{-2}$  &0.002& 0.8 &$0.018$ & $0.17$   &3.10 & 5.18 &$2048\times 256^2$ \\
    \hline 
 \end{tabular}
 \caption{\label{table} Parameters of the numerical simulations.
   $\Rep=U_c\,d/\nu$, $\Re = u_\mathrm{rms}L/ \nu$: Reynolds number,
   $u_\mathrm{rms}$: root-mean-square velocity, $\varepsilon$:
   mean kinetic energy dissipation rate, $\nu$: kinematic viscosity,
   $d$: sphere diameter, $\eta =(\nu^3/\varepsilon)^{1/4}$:
   Kolmogorov dissipation length scale, $\tau_\eta =
   (\nu/\varepsilon)^{1/2}$: Kolmogorov time scale, $L =
   u_\mathrm{rms}^{3}/\varepsilon$: integral scale, $T_L =
   L/u_\mathrm{rms}$: large-eddy turnover time, $N^3$: number of
   collocation points.}
\end{table*}
\endgroup

The physical configuration is determined by three dimensionless
parameters: the particle Reynolds number $\Rep$, the Reynolds number
of the fluid $\Re$, and the turbulent large-scale intensity
$I=u_\mathrm{rms}/U_c$. The latter measuring the strength of the large
scale turbulent fluctuations compared to the mean flow velocity. For
each particle Reynolds number $\Rep$ considered we varied the
turbulent intensity $I$ between 0.05 and 0.60. Although the parameter
space is three-dimensional we vary only two parameters namely $\Rep$
and $I$. This is due to our particular choice of keeping constant, for
each value of $\Rep$, the mean velocity $U_c$, the viscosity $\nu$,
the particle diameter $d$ and the form of the external forcing but to
vary only its amplitude. This latter control parameter determines in
turn the root-mean-square velocity $u_\mathrm{rms}$, which enters in
the definitions of both $I$ and $\Re$.

\label{sec:turb}
\begin{figure}
  \begin{center}
    \includegraphics[width=0.49\textwidth]{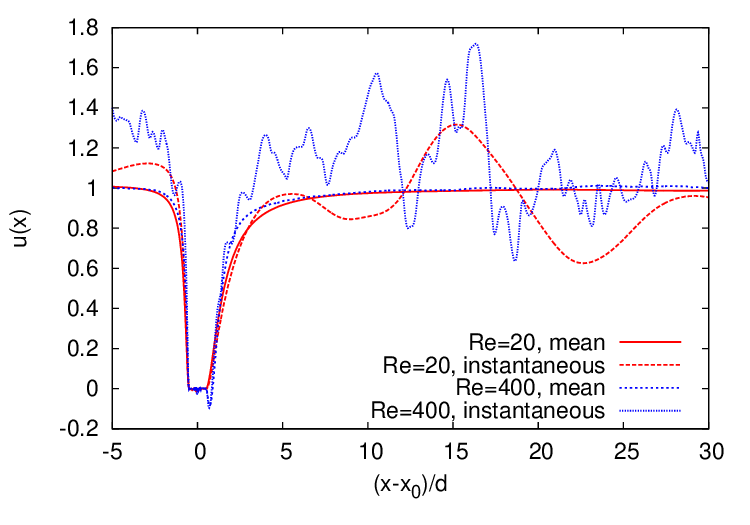}
    \includegraphics[width=0.49\textwidth]{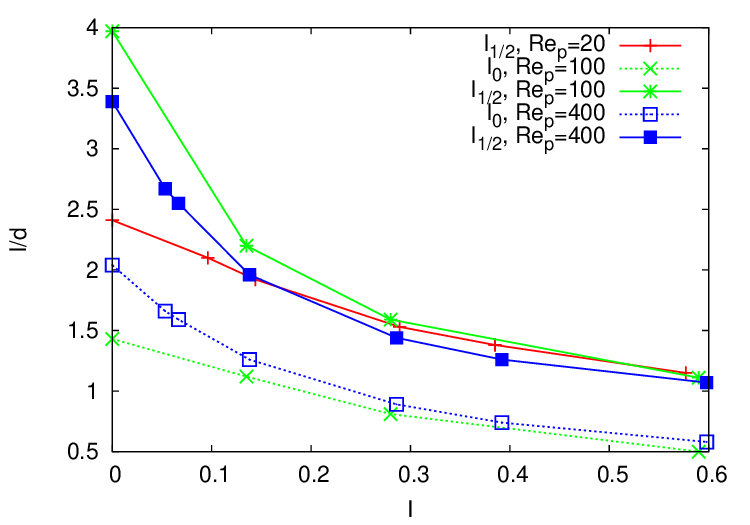}
  \end{center}
  \caption{\label{fig:profile_20_400} Left: Instantaneous velocity
    profile, together with its temporal average; the sphere center is
    here located at $x_0$.  Right: averaged
    velocity profiles for $\Rep=400$ and various values of the
    turbulent intensity $I$ as labelled.}
\end{figure}
A first qualitative idea of the flow structure around the sphere can
be observed from Fig.~\ref{fig:profile_20_400} (Left) where
instantaneous velocity profiles along the axis of symmetry, together
with their temporal averages, are shown for two particle Reynolds
numbers ($\Rep=20$ and $\Rep=400$) in the case of a turbulent
intensity $I\approx 0.25$. From the instantaneous profiles we remark
that, for a given turbulent intensity, the higher is the particle
Reynolds number, the smaller are the typical lengthscales of the
turbulent fluctuations. Also, when fixing $\Rep=400$ and varying the
intensity $I$, we see in Fig.~\ref{fig:profile_20_400} (Right) that
stronger turbulent fluctuations reduce the extent of the mean particle
wake. This confirms observations made by
\cite{bagchi-balachandar:2004}.

\subsection{The drag and lift forces}
\begin{figure}
  \begin{center}
    \includegraphics[width=0.49\textwidth]{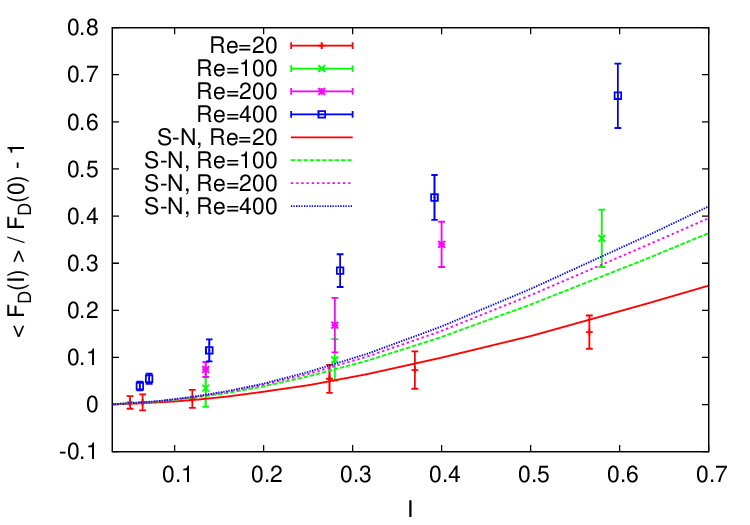} \hfill
    \includegraphics[width=0.49\textwidth]{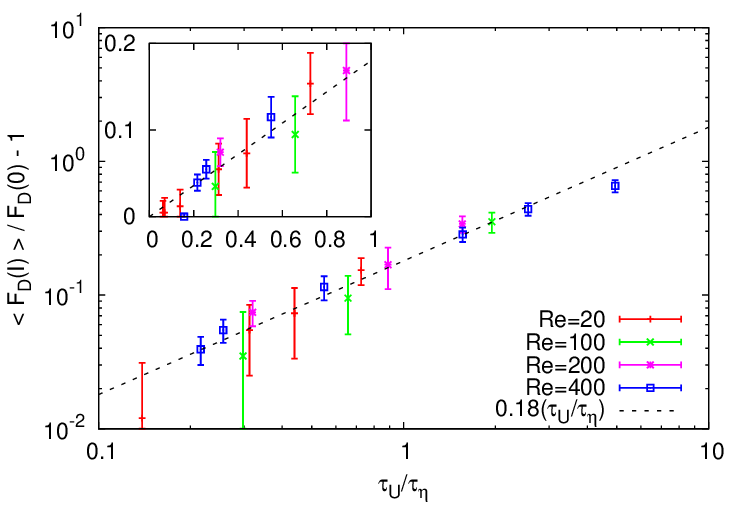}
  \end{center}
  \caption{\label{fig:meanDrag} Left: deviations of the mean drag from
    that obtained with a laminar upstream flow as a function of the
    turbulent intensity $I$. Right: same represented as a function of
    the ratio $\tau_\mathrm{U}/\tau_\eta$ where $\tau_\mathrm{U} =
    d/U_c$ is the sweeping time over a distance equal to the particle
    diameter and $\tau_\eta = (\nu/\varepsilon)^{1/2}$ is the
    turbulent turnover time associated with the Kolmogorov dissipative
    scale $\eta$.}
\end{figure}

We computed the mean drag, i.e.\ the force exerted by the turbulent
flow on the particle in the streamwise direction (the averaged
perpendicular forces are zero). The total force acting on the particle
also contains the volume forcing applied to maintain the carrier flow
in a stationary turbulent state.  However, we have estimated this
force and could show that its contribution is negligible compared to
the drag and lift forces exerted by the fluid.  Results are shown in
Fig.\ref{fig:meanDrag} (Left), where data is normalised to the case
without turbulent fluctuations. The error bars are given by measuring
the standard deviation from the mean and estimating how many
independent realisations our temporal signal contains. For that we
approximate the correlation time by the zero-crossing time of the
auto-correlation functions of the drag forces (see later).
One finds a clear increase of the stream-wise force on the particle
with increasing turbulent intensity. For instance, when $\Rep=400$ and
$I = 0.5$, we measure a relative increase of 40\%.  Indications for
this increased drag were already given by \cite{Bagchi2003} and
\cite{Kim2011}. However, their turbulent intensities were too small
and the error-bars were too large to clearly distinguish their
measured drag from the drag obtained without turbulence and validate
undoubtedly the observed increase.

As stated in \cite{Bagchi2003}, it is very likely that the drag
increase is due to the non-linear dependence of the drag upon the
incident fluid velocity. The instantaneous slip velocity of the
particle is given by $\bm U = U_c\,\bm e_x +\mathcal{U}$ where the
large-scale turbulent velocity $\mathcal{U}$ can be approximated by a
Gaussian random vector with zero mean and variance
$u_\mathrm{rms}^2$. The fluctuations of $\mathcal{U}$ can be
responsible for an increase of the force $\bm F$. Indeed, let us
assume that the relation between $\bm F$ depends and $\bm U$ is
instantaneously given by the Schiller and Naumann approximation
formula (\ref{eq:schillerNaumann}), that is
\begin{equation}
  \bm F \approx \bm F^{\mathrm{SN}}(\bm U)= 3\pi\,d\,\nu \left[1+0.15
  \left(\frac{|\bm{U}|\,d}{\nu} \right)^{\!0.687}\right]\,\bm U.
\end{equation}
The average of this force with respect to the symmetric Gaussian
fluctuations of $\mathcal{U}$ is aligned with the streamwise direction
$x$ and reads
\begin{eqnarray}
\overline{\bm F}^{\mathrm{SN}}(I) &=&
\frac{1}{\left(\sqrt{2\pi}\,u_\mathrm{rms}\right)^{3}} \int \bm
F^{\mathrm{SN}}(\bm U)\,
\mathrm{e}^{-|\mathcal{U}|^2/(2u_\mathrm{rms}^2)}\,\mathrm{d}^3\mathcal{U}
\nonumber\\  &=& 3\pi\,d\,\nu\,U_c\,\bm
e_x\left[1+\frac{0.15\,\Rep^{0.687}}{(2\pi)^{3/2}} \!\!\int (1 \!+\! I\,v_1) |\bm
    e_x \!+\! I\,\bm v|^{0.687}\,\mathrm{e}^{-|\bm
      v|^2/2}\,\mathrm{d}^3v\right]\!\!.
  \label{eq:fluctu_fsn}
\end{eqnarray}
Although the integral cannot be written explicitly, one can easily
check that it leads to a relative increase of the drag that takes of
the form $\Delta^{\mathrm{SN}}(I) =
\overline{F}^{\mathrm{SN}}(I)/\overline{F}^{\mathrm{SN}}(0)-1 \simeq
C\,I^2$ where $C\approx 0.45\,\Rep^{0.687}/(1+0.15\,\Rep^{0.687})>0$.
Also, the estimate (\ref{eq:fluctu_fsn}) does surprisingly not depend
on the fluid Reynolds number $\Re$ but only on the two other
dimensionless parameters $I$ and $\Rep$.  The prediction
(\ref{eq:fluctu_fsn}) is shown as lines in Fig.~\ref{fig:meanDrag}
(Left). While it gives a rather good approximation of the data at
moderate particle Reynolds numbers, it is clearly a too low estimate
for $\Rep > 100$. The discrepancy might be due to the impossibility
for this approach to account for the dependence on $\Re$.

Another heuristic way to understand the effect of the perturbing
turbulence on the drag consists in considering the modification of the
amplitude of the typical velocity gradient in the neighbourhood of the
particle.  In the laminar case (when $I=0$), the velocity field around
the particle typically varies over a scale of the order of $d$. The
unperturbed gradient is then $\sim U_c/d$. When a background
turbulence is added to this flow, the value of the gradient is
modified by the typical turbulent gradient $u_\eta/\eta =
\tau_\eta^{-1} = \nu^{-1/2}\varepsilon^{1/2}$ where $\tau_\eta$ is the
turnover time associated to the Kolmogorov dissipative scale
$\eta$. The velocity gradients appear in the viscous forces acting on
the particle and also at leading order on the pressure gradient via
the Poisson equation. Altogether these considerations imply that the
corrections to the drag are expected to be $\propto \tau_\eta^{-1}$,
while these forces in the $I=0$ reference case should be $\propto
\tau_U^{-1}$, where $\tau_U = d/U_c$. Such arguments leads to a
relative increase $\Delta(I)$ of the drag force acting on the particle
$\propto \tau_U/\tau_\eta = I^2 \Rep\,\Re^{-1/2}$. The right-hand side
of Fig.~\ref{fig:meanDrag} shows $\Delta(I)$ as a function of
$\tau_U/\tau_\eta$ and reveals a rather good collapse of the entire
dataset to the curve $\Delta(I) = 0.18\,(\tau_U/\tau_\eta)$, giving
support to such phenomenological arguments. Notice that this approach
gives again a behaviour $\propto I^2$ for small values of the
turbulence intensity, but with this time a constant that depends
explicitly on the outer-flow Reynolds number $\Re$.

There are thus at least two different mechanisms that explain the drag
increase by the impacting turbulence: The non-linear dependence of the
drag on the incident large-scale velocity and the modification of the
particle boundary layer by small-scale turbulent fluctuations. These
two effects might act concomitantly at small turbulent
intensities. However, data show that the drag increase strongly
depends on the particle Reynolds number. This effect is not captured
by the first approach. In addition the collapse of data in
Fig.~\ref{fig:meanDrag} (Right) gives support to the importance of
shear effects. Nevertheless, one cannot firmly favour one of these two
mechanisms as the qualitative consequences of an outer turbulence
depend on the particle Reynolds number.

\begin{figure}
  \begin{center}
    \includegraphics[width=0.49\textwidth]{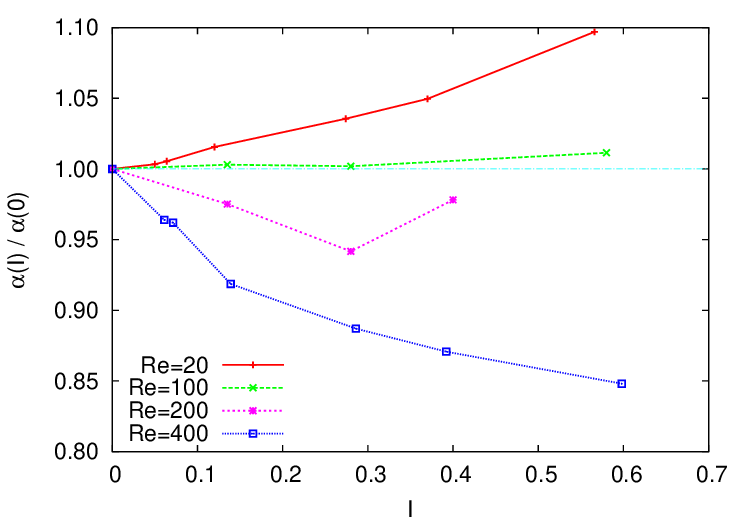} \hfill
    \includegraphics[width=0.49\textwidth]{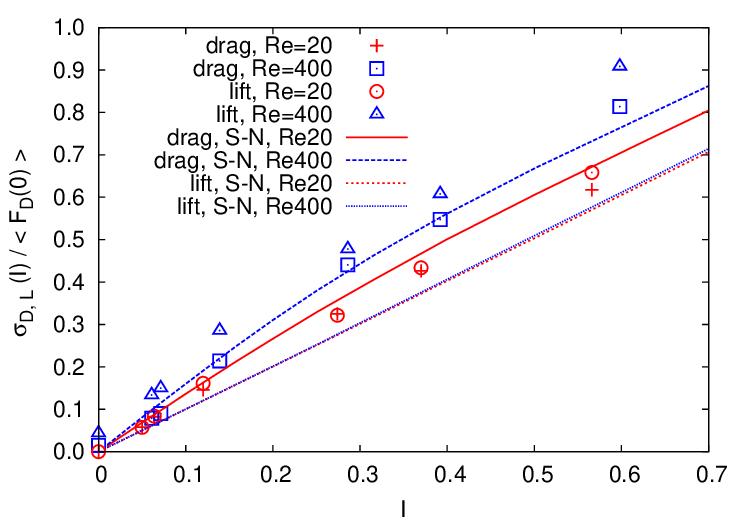}
  \end{center}
  \caption{\label{fig:meanDrag_std} Left: normalised ratio $\alpha(I)
    = \langle F_p(I) \rangle / \langle F_v(I) \rangle $ of pressure to
    viscous contributions to the drag force. Right: standard deviation
    of the drag force $\sigma_D(I) = \langle [F_D(I) - \langle
    F_D(I)\rangle]^2\rangle^{1/2}$ and of the lift force $\sigma_L(I)
    = \langle [F_L(I) - \langle F_L(I)\rangle]^2\rangle^{1/2}$ as a
    function of the turbulent intensity. In each case it is normalized
    to the mean value of the drag force for zero turbulent
    intensity. The solid lines correspond to the prediction obtained
    from a fluctuating Schiller--Naumann relation.}
\end{figure}

An instance of differing behaviours depending on $\Rep$ is the ratio
between the respective contributions $F_p$ and $F_v$ of pressure and
viscosity to the drag. We show in the left-hand side of
Fig.~\ref{fig:meanDrag_std} their normalised ratio $\alpha(I) =
\langle F_p(I) \rangle / \langle F_v(I) \rangle $ for different
turbulent intensities. For the lowest particle Reynolds number
($\Rep=20$), the pressure contribution to the total force increases
with the turbulent intensity while one observes the opposite for the
highest Reynolds number ($\Rep=400$). At the intermediate
configuration with $\Rep=100$ this ratio is more or less independent
of the turbulent intensity. Recall that when $\Rep\lesssim 100$ the
viscosity contribution is dominant while when $\Rep\gg 1$ the pressure
largely dominates the drag force. The viscosity and pressure give
approximately equal contributions for a particle Reynolds number of
100 (see also the inset of Fig.~\ref{fig:re_drag_coeff}). It seems
from our data that the presence of turbulent fluctuations
counterbalances these pressure-viscosity ratios which depend on
$\Rep$. We will see in the following section that the reason for this
can be found in the modification of the flow structure around the
sphere by the fluctuating nature of the ambient flow at small scales.

After having examined the average forces we now turn to the
fluctuations of the drag and lift forces. Their amplitudes can be
characterised by their standard deviation, which is shown in the
right-hand side of Fig.~\ref{fig:meanDrag_std}. We observe that the
standard deviation increases with increasing turbulent intensity. The
fluctuations of the lift forces are slightly larger than those of the
drag, as also observed by \cite{Bagchi2003}. As the upstream flow is
isotropic this difference can only be attributed to the boundary layer
of the sphere and its wake. Using the approach described above we used
Schiller--Naumann relation for the drag to derive a formula equivalent
to (\ref{eq:fluctu_fsn}) but for the force standard deviation.  The
resulting predictions are plotted as solid lines in
Fig.~\ref{fig:meanDrag_std} (Right). Interestingly, this approach
gives a reasonable prediction, indicating that the force fluctuations
are mainly due to large-scale fluctuations. We will come back to this
conclusion in the following section.

\begin{figure}
  \begin{center}
    \includegraphics[width=0.49\textwidth]{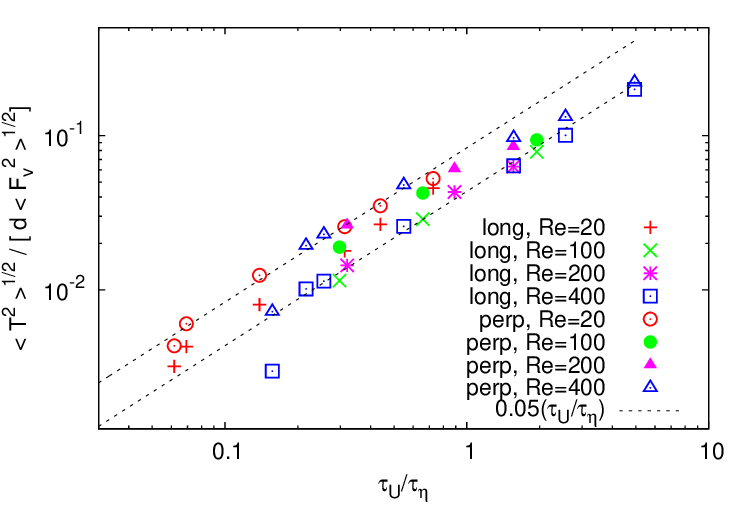}
    \hfill
    \includegraphics[width=0.49\textwidth]{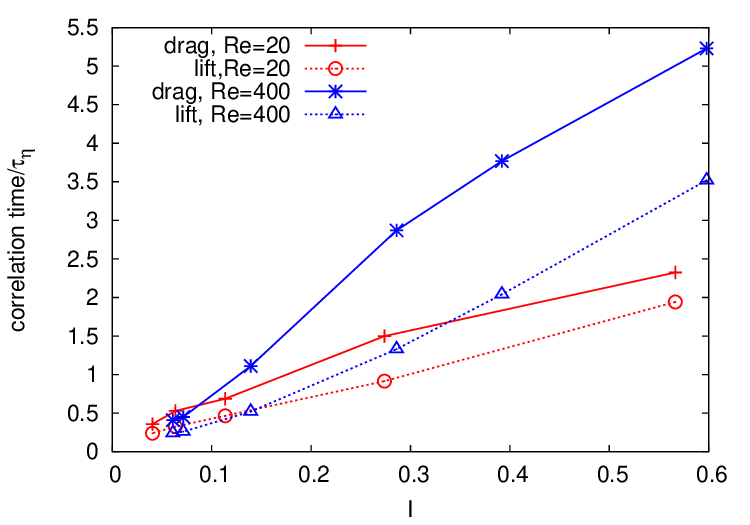}
  \end{center}
  \caption{\label{fig:torque} Left: standard deviation of the
    component of the torque exerted by the flow onto the particle as a
    function of the ratio $\tau_U/\tau_\eta$. The streamwise (long)
    and the spanwise (perp) components show different
    behaviours. Right: correlation times of the drag and lift forces
    (see text for definition).}
\end{figure}
The measured standard deviation of the torque acting on the particle
(\ref{eq:torque}) is shown in Fig.~\ref{fig:torque}. There we have
normalised it to what we expect to be its typical value, namely the
typical amplitude of the viscous force multiplied by the particle
diameter $d$. Interestingly, we observe an increase of the torque as a
function of the ratio of gradients $\tau_U/\tau_\eta$. The datapoints
collapse to a linear behaviour as a function of $\tau_U/\tau_\eta$,
with a constant that depends whether streamwise or span-wise
fluctuations are considered. As for the average drag, the explanation
of this linear behaviour relies on the fact that the fluctuations of
viscous forces are directly related to the turbulent velocity
gradients that surround the particle. We mention that the torque in
the stream-wise direction is always smaller than in the transverse
directions. Let us also stress that the particle is not allowed to
rotate.

We now turn to the two-time statistics of the force exerted by the
fluid on the particle by measuring the temporal correlations
$C(t)=\langle F(t)\,F(0) \rangle $ of the drag $F_D$ and of the lift
$F_L$.  The correlations of the drag always last longer than those of
the lift. This property has also been observed in \cite{Kim2011} and
can be interpreted by the fact that the drag is more sensitive to
large-scale fluctuations, while the lift samples shorter
timescales. To quantify this effect, we introduce the correlation time
$\tau_c$ as the time needed to half the correlation of the forces,
i.e.\ $C(\tau_c) = (1/2)\,C(0)$. The behaviour of this correlation
time as a function of $I$ is shown in the right-hand side of
Fig.~\ref{fig:torque} for $\Rep=20$ and $\Rep=400$. This graph shows
clearly the increase mentioned above.

\begin{figure}
  \begin{center}
    \includegraphics[width=0.49\textwidth]{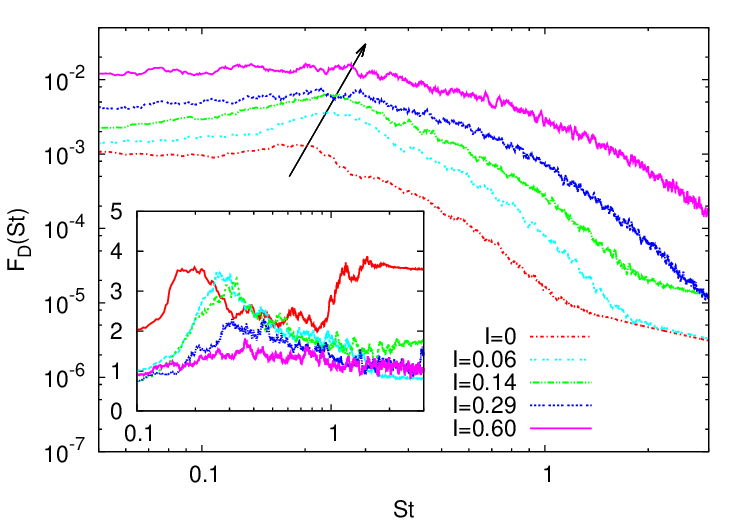}
    \hfill
    \includegraphics[width=0.49\textwidth]{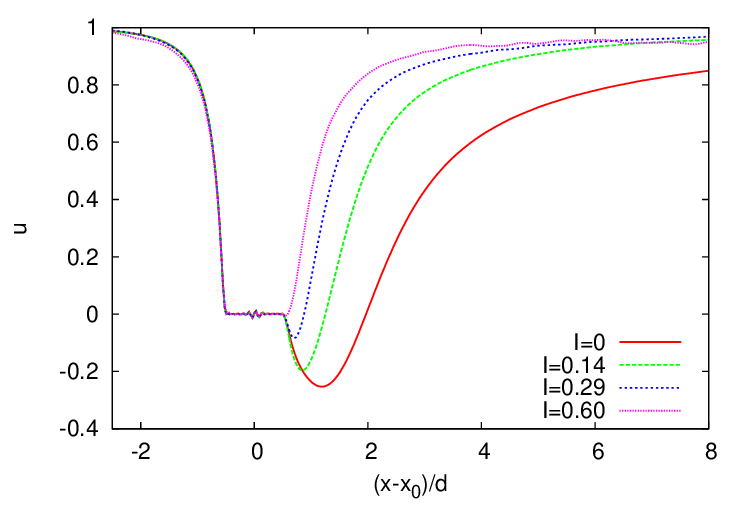} 
  \end{center}
  \caption{\label{fig:freq} Left: frequency spectrum of the lift
    forces on the particle as a function of the Strouhal number
    $\St=f\, d/U_c$. Inset: ratio of the lift to the drag
    amplitude. Right: half-wake length $l_{1/2}$ defined as
    $u(l_{1/2})=1/2 U_c$ and size of the recirculation region $l_0$
    defined as $u(l_0)=0$.}
\end{figure}

We also present measurements on the frequency spectrum of the drag and
lift forces. For the low-$\Rep$ case no difference can be observed
between the drag and lift components (not shown). For $\Rep=400$, a
couple of remarkable features can be observed in Fig.~\ref{fig:freq} (Left),
which shows the amplitudes of the forces Fourier modes as a function
of the Strouhal number $\St=f\, d/U_c$, where $f$ is the frequency. In
order to reduce the statistical noise we applied a window average.  At
$\Rep=400$ the wake is intrinsically turbulent and we observe a peak
in the spectrum at at $\St^\star=0.2$ for the lift force at $I=0$,
which matches the eddy-detachment frequency of the wake. Such a peak
is not visible for the drag amplitude (not shown). It can be more
clearly seen from the inset which shows the ratio of the lift and the
drag amplitudes. Two observations deserve attention: first, an
increasing turbulent intensity slightly increases this detachment
frequency, thus $St^\star=St^\star(I)$. Secondly, the signature of
this frequency survives up to a turbulent intensity of at least
0.29. This is surprising as the amplitude of the turbulent
fluctuations at this intensity is larger than the peak-value
$F_D(St^\star)$ of a laminar inflow. This means that the turbulent
fluctuations increase the violence with which vortex are detached in
the wake. This can be related to the increase of the gradients in the
mean velocity profile and thus to the shortening of the particle wake
that is studied with more details in next section.

\subsection{Average velocity boundary layer and wake}

As already stated, the particle wake is shortened by the presence of
upstream turbulent fluctuations. To quantify this, we measure the
half-wake length $l_{1/2}$, which is the distance from the centre of
the sphere of the location where the average velocity deficit on the
symmetry axis is equal to one half. Data is shown in
Fig.~\ref{fig:freq} (Right). Already a turbulent intensity of 0.3-0.4
is sufficient in both cases to divide by two the size of the laminar
wake. Notice that these wake reductions can have important
consequences for the interaction between many particles in
particle-laden problems, such as sedimentation. We also note that for
turbulent intensities larger than 0.2 the wake size normalised by $d$
reaches a behaviour that is identical for all the particle Reynolds
numbers considered.

The observed steepenings of the mean velocity gradients in the
downstream neighbourhood of the particle are consistent with the
results reported previously on the increase of the drag force. Such
sharp gradients increase the stress, which leads in turn to an
increased viscous drag force ---\,see (\ref{eq:forceParticle}). The
average pressure profile around the sphere is also modified by
stronger average velocity gradients. The down-stream surface pressure
decreases as a function of the turbulent intensity while the up-stream
pressure remains almost unchanged (not shown). Thus, the pressure
contribution to the drag force also increases with the turbulent
intensity. However, we remark that the ratio of these contributions
depends on the turbulent intensity (see the left-hand side of
Fig.~\ref{fig:meanDrag_std}).

To understand more quantitatively the modifications of the wake
by the turbulent perturbations when $I\ll1$, let us decompose the
velocity field as $\bm u = \bm u_0 +\bm u'$, where $\bm u_0$ is the
velocity obtained in the case of a laminar upstream flow (i.e.\ when
$I=0$), which contains the possible tubulent wake of the particle, and
$\bm u'$ denotes the perturbation due to inflow turbulence (i.e.\ $\bm
u'=0$ when $I=0$). When averaging with respect to the turbulent
fluctuations, the mean velocity profile will be the sum of the
unpertubed mean profile $\bm U_0 = \overline{\bm u_0}$ and of the
averaged perturbation $\bm U' = \overline{\bm u'}$, which vanishes
only far away from the particle. One can easily check writing the
Reynolds-averaged equation that, to leading order in $I$, these
velocities satisfy
\begin{equation}
  \bm U' \cdot \nabla  \bm U_0 + \bm U_0 \cdot \nabla \bm U' - \nu
  \nabla^2 \bm U'= -\nabla\cdot\bm\tau' + \nabla
  P',\quad\nabla\cdot\bm U' = 0,
  \label{eq:rans2}
\end{equation}
where $\bm\tau'$ is the Reynolds stress tensor defined as $\tau_{ij} =
\overline{u_i'\,u_j'}$ and $P'$ is the correction to the mean pressure
due to the presence of upstream turbulence. As we see from
(\ref{eq:rans2}) the perturbation $\bm U'$ of the mean velocity profile
originates from the Reynolds stress. The pressure just appears to
maintain the divergence-free condition.

\begin{figure}
  \begin{center}
    \includegraphics[width=0.49\textwidth]{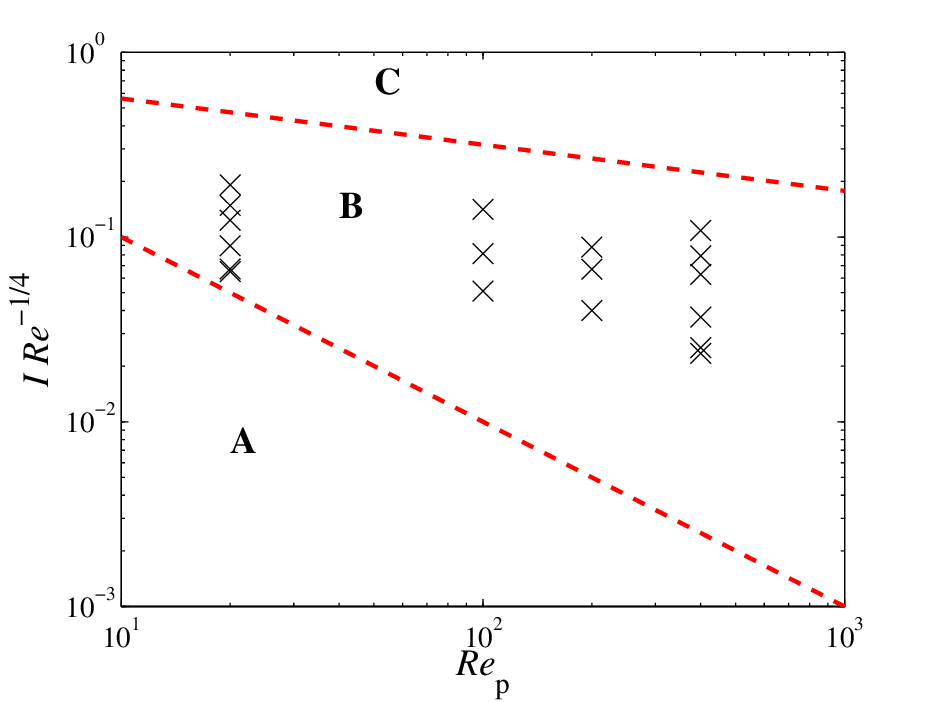}
    \hfill
    \includegraphics[width=0.49\textwidth]{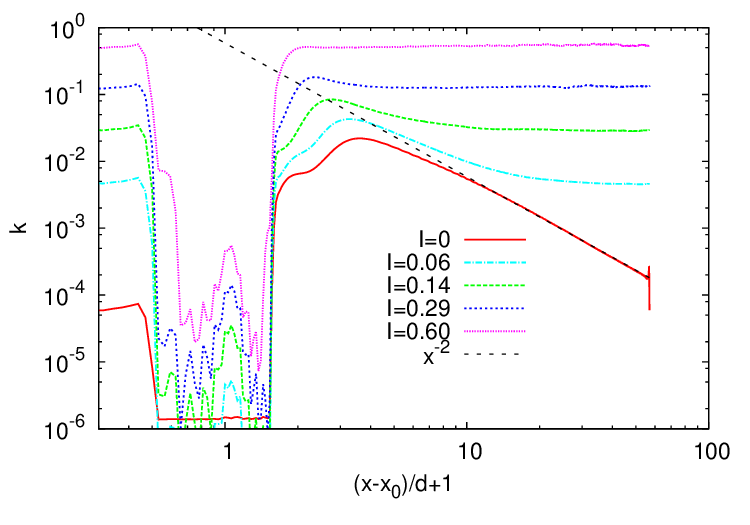}
  \end{center}
  \caption{\label{fig:sketch_regimes} Left: representation in the parameter plane
    $(\Rep,I\,\Re^{-1/4})$ of the three regimes for the wake disturbance (see text). The
    black crosses are the values corresponding to the numerical
    simulations. Right: turbulent kinetic energy profile along the
    symmetry axis for $\Rep=400$.}
\end{figure}

Let us estimate the order of magnitude of the different terms
appearing in (\ref{eq:rans2}). It is clear that $\bm U_0$ is of the
order of the mean upstream velocity $U_c$ and that it varies on scales
of the order of the particle diameter $d$. The mean perturbation $\bm
U'$ has an unknown amplitude $U'$ that we want to characterise.  Also,
$\bm U'$ varies over a scale $\ell$ that is different from $d$. To
estimate this scale, let us do the following observation.  We expect
all the turbulent eddies of scales $r\lesssim d$ to be affected in the
vicinity of the particle by its presence.  While they are swept
downstream the particle, we expect the eddies of size $r$ to recover
their usual turbulent characteristics after a time of the order of
their correlation time $\tau_r$.  Kolmogorov's 1941 scaling gives for
$r$ inside the inertial range $\tau_r\sim\varepsilon^{-1/3} r^{2/3}$,
where $\varepsilon$ is the turbulent kinetic energy dissipation
rate. The scale $\ell$ of variation of the perturbed wake $U'$ is
given by the distance downstream the particle that is travelled by the
largest perturbed eddies before forgetting they have met the
particle. One then distinguishes two cases depending whether $d$ is
larger or smaller than the dissipative scale $\eta$ of upstream
turbulence. With such phenomenological
considerations, 
\begin{itemize}
\item When $d\gg\eta$, we have $\ell\sim U_c\, \tau_d \sim U_c\,
  \varepsilon^{-1/3} d^{2/3}$, so that $\ell/d \sim
  I^{-4/3}\Re^{1/3}\Rep^{-1/3}$. One can then estimate the magnitude
  of the terms present in the left-hand side of (\ref{eq:rans2}):
  \begin{eqnarray}
    \bm U' \cdot \nabla \bm U_0 &\sim& U'\,U_c/d, \nonumber \\
    \bm U_0 \cdot \nabla \bm U' &\sim& U_c\,U'/\ell \sim
    (U'\,U_c/d) (d/\ell)\sim
    (U'\,U_c/d)\,I^{4/3}\Re^{-1/3}\Rep^{1/3},\nonumber\\
    \nu \nabla^2 \bm U' &\sim& \nu\,U'/\ell^2 \sim (U'\,U_c/d)
    (d/\ell)^2 \Rep^{-1} \sim
    (U'\,U_c/d)\,I^{8/3}\Re^{-2/3}\Rep^{-1/3},\nonumber\\
    \nabla\cdot\bm\tau&\sim& u_\mathrm{rms}^2/\ell \sim (U_c^2/d)\,
    I^{10/3}\Re^{-1/3}\Rep^{1/3}.
    \label{eq:dimestimate}
  \end{eqnarray}
  For $\Re\gg1$, $\Rep\gg1$ and $I\lesssim1$, one always have
  $I\,\Re^{-1/4}\ll\Re_p^{1/2}$, so that the viscous term is
  negligible compared to $\bm U_0 \cdot \nabla \bm U'$.  One of the
  two first quantities thus has to balance the turbulent Reynolds
  stress.
\item When $d\ll\eta$, that is $I\,\Re^{-1/4}\ll\Re_p^{-1}$, the
  situation is different: at the scale of the particle, the turbulent
  flow approximately appears as a uniform gradient with a correlation
  time equal to the Kolomogorov scale turnover time $\tau_\eta =
  \nu^{1/2}\varepsilon^{-1/2}$. The scale of variation of the wake
  perturbation is thus $\ell \sim U_c\,\tau_\eta$, so that $\ell/d \sim
  I^{-4/3}\Re^{1/2}\Rep^{-1}$ in that case. The estimation of the various
  terms of (\ref{eq:rans2}) are then
\begin{eqnarray}
  \bm U' \cdot \nabla \bm U_0 &\sim& U'\,U_c/d, \nonumber \\
  \bm U_0 \cdot \nabla \bm U' &\sim&
   (U'\,U_c/d)\,I^{4/3}\Re^{-1/2}\Rep,\nonumber\\ 
  \nu \nabla^2 \bm U' &\sim&  (U'\,U_c/d)\,I^{8/3}\Re^{-1},\nonumber\\
  \nabla\cdot\bm\tau&\sim& u_\mathrm{rms}^2/\ell \sim  (U_c^2/d)\, I^{3}\Re^{-1/4}\Rep.
\end{eqnarray}
Again in that case, the viscous term is always subdominant. One can
also easily check that the term $\bm U' \cdot \nabla \bm U_0$ is
always dominant and thus has to balance the Reynolds stress.
\end{itemize}
Finally, these two conditions and the balance between the various
terms lead to distinguish between three different cases:
\begin{description}
\item[A:\ ] $I\,\Re^{-1/4}\ll\Rep^{-1}$. This corresponds to parameter
  values such that $d\ll\eta$. The dominant effect that compensates
  Reynolds stress is the stretching of the perturbation $\bm U'$ by
  the mean flow $\bm U_0$. The length of the wake perturbation is
  $\ell/d \sim I^{-4/3}\Re^{1/2}\Rep^{-1}\gg 1$ and its amplitude is
  $U'/U_c \sim I^4\,\Re^{-1/2}\,\Rep$.
\item[B:\ ] $\Rep^{-1}\ll I\,\Re^{-1/4}\ll\Rep^{-1/4}$.  This
  corresponds now to $d\gg\eta$. The dominant effect is again the
  stretching of the perturbation $\bm U'$ by the mean flow $\bm
  U_0$. However the length of the wake perturbation is now $\ell/d \sim
  I^{-4/3}\Re^{1/3}\Rep^{-1/3}\gg 1$ and its amplitude is $U'/U_c \sim
  I^{10/3} \Re^{-1/3} \Rep^{1/3}$.
\item[C:\ ] $I\,\Re^{-1/4}\gg\Rep^{-1/4}$. We still have here $d\gg\eta$
  but the dominant effect is now the advection of the perturbation
  $\bm U'$ by the mean flow $\bm U_0$. The length of the wake
  perturbation is again $\ell/d \sim I^{-4/3}\Re^{1/3}\Rep^{-1/3}\ll 1$
  but its amplitude is now $U'/U_c \sim I^{2}$ and is thus independent
  on $\Re$ and $\Rep$.
\end{description}
Surprisingly the transitions between these regimes depends on only two
parameters: the particle Reynolds number $\Rep$ and the product
$I\,\Re^{-1/4}$. These three regimes are represented in the plane of
these two parameters in Fig.~\ref{fig:sketch_regimes} (Left), together with
the parameters of the present numerical study. All datapoints are in
the region \textbf{B} of the parameter space. This can be explained by
the fact that in a turbulent state, for a fixed small intensity $I$,
being in regime \textbf{A} would require a large value of $\Re$ and in
\textbf{C}, a large value of $\Rep$.  In both cases, this would
increase significantly the required resolution beyond reasonable
numerical means.

In the regime \textbf{B}, the scale $\ell$ of the perturbation to the
average velocity profile is much larger than the particle diameter
$d$. This implies that the whole wake is perturbed, including the part
of it that is the nearest to the particle. Also, in this regime, the
normalised perturbation of the mean velocity gradient has the order of
magnitude $(U'/\ell) / (U_c/d) \sim I^{14/3} \Re^{-2/3} \Rep^{2/3}$,
which grows very fast as a function of the turbulence intensity $I$.
These arguments seem to confirm our speculation on the small-scale
modifications of the boundary layer by the turbulent fluctuations to
cause the increase of the average drag force. However,
\cite{merle-legendre-etal:2005} showed that the whole range of
turbulent scales contribute to the forces on a bubble.  It is thus
possible that the characteristic scale $\ell$ that we introduced above
actually spread over the full inertial range. Confirming its existence
would require very-fine numerical measurements that are beyond the
scope of the present work.

To complete these results, let us add some comments on the observed
behaviour upstream and downstream the particle. The upstream profile
does not seem much altered by the turbulent fluctuations whereas the
wake looks heavily modified (see Fig.~\ref{fig:profile_20_400}
Right). Upstream we find a clear power-law for the dependence of the
velocity deficit as a function of the distance to the particle
centre. The measured exponent varies from $\approx -2.5$ for $\Rep=20$
to $\approx -3$ for $\Rep=400$ and seems independent of the turbulent
intensity. The gradients in the boundary layer introduced by the
turbulent carrier flow are still subdominant in the upstream region
and thus do not change the upstream boundary layer on average. On the
other side, the downstream profiles change with both the particle
Reynolds number and the turbulent intensity. However, we always find a
velocity deficit scaling $\propto x^{-1}$ for the large Reynolds
number, independently of $I$. Also, as seen in
Fig.~\ref{fig:sketch_regimes} (Right), we observe a $x^{-2}$
intermediate decay of the turbulent kinetic energy that is absorbed by
outter turbulence when $I$ increases. We do not observe the $x^{-2}$
law for the velocity deficit reported by
\cite{amoura-roig-etal:2010}. This law should appear as an
intermediate asymptotics \citep{eames-johnson-etal:2011} that we might
not resolve.

\section{Conclusion}
\label{sec:conclusion}

In this article we address the question how turbulent fluctuations in
the carrier flow influence the drag and lift forces acting on a towed
sphere.  We find that the average drag significantly increases
compared to what is know for a laminar upstream flow. Also the torque
of the particle increases as a function of the turbulent
intensity. These increases depend on the ratio of the flow advection
time across the particle and the Kolmogorov time scale thus assumed to
be a small-scale effect. The fluctuations of the drag force are
satisfactorily described by standard drag correlations such as that of
\cite{schiller-naumann:1933}. Turbulent fluctuations in the upstream
flow increase the frequency and strength of the wake detachment at
$Re_p=400$. The implications of turbulence on the boundary layer and
wake of the particle are also addressed. This work also primarily
aimed at validating a new method for integrating the incompressible
Navier--Stokes equation with no-slip boundary conditions on the
surface of a moving particle. Simulations that were performed with
this method in the case of a spherical particle embedded in a laminar
upstream flow, show that such a strategy reproduces with a high
accuracy known results on the variation of the drag and lift forces as
a function of the particle Reynolds number and give a precise handling
on the turbulent fluctuations that arise in the wake downstream the
sphere.

Such benchmarking will now allow us confidently attacking the physical
problems that motivated the development of the code. The first
question that is subject to ongoing work concerns the dynamics of
neutrally buoyant finite-size particles that are transported by a
developed turbulent flow in the case when the diameter of such
particles falls inside the turbulent inertial range of length
scales. The drag, lift, and the slip velocity associated to such
settings are still not well defined because of the fluctuating nature
of the fluid surrounding environment. This situation still requires an
important modelling effort. Today's models usually rely on applying to
turbulent situations formula that are known only for upstream laminar
flows, as e.g.\ \cite{schiller-naumann:1933}. Finite-size models can
now be improved with the help of the present results. This work will
also give new insights for understanding how the fluid velocity
statistical properties are modified by the presence of the
particle. Preliminary results on a moving particle confirm that a
turbulent boundary layer develops around the particle and that its
thickness is comparable to the particle size.

Another important open question that is worth mentioning here relates
to possible improvements of the order of the method that is presented
here. As we have seen, the error of the proposed method is of the
order of $\Delta x^{3/2}$. To go beyond this first order, a natural
idea is to make use of the variational formulation that is described
in Sec.~\ref{sec:method} and which consists in expressing the penalty
force as a solution to a minimisation problem. To avoid Gibbs
oscillations that, as we showed, are responsible for the observed
error, it could be envisaged to relax such minimality to the surface
instead of the full particle volume, allowing for instance the fluid
flow to develop inside the particle, without affecting the no-slip
boundary condition. Such ideas lead to new difficulties related to the
non-locality of pressure and which still need to be cautiously worked
out.

\subsection*{Acknowledgements}
\noindent This work benefited from fruitful discussions with
J. Dreher. Part of this research was supported by the French Agence
Nationale de la Recherche under grant No.\ BLAN07-1\_192604. The
research leading to these results has received funding from the
European Research Council under the European Community's Seventh
Framework Program (FP7/2007-2013, Grant Agreement no. 240579). Access
to the IBM BlueGene/P computer JUGENE at the FZ J\"ulich was made
available through project HBO22. Part of the computations were
performed on the “m\'esocentre de calcul SIGAMM” and using HPC
resources from GENCI-IDRIS (Grant 2009-i2010026174).

\bibliographystyle{jfm.bst}

\end{document}